\documentclass[a4paper,UKenglish,cleveref, autoref, thm-restate]{lipics-v2021}

\title{Prefix Sorting DFAs: a Recursive Algorithm} 

\author{Nicola Cotumaccio}{GSSI, Italy \and Dalhousie University, Canada}{nicola.cotumaccio@gssi.it}{https://orcid.org/0000-0002-1402-5298}{}

\authorrunning{N. Cotumaccio} 

\Copyright{Nicola Cotumaccio} 

\ccsdesc[500]{Theory of computation~Graph algorithms analysis}
\ccsdesc[500]{Theory of computation~Pattern matching}

\keywords{Suffix Array, Burrows-Wheeler Transform, FM-index, Recursive Algorithms, Graph Theory, Pattern Matching.} 

\category{} 

\funding{This work was partially funded by Dante Labs.}

\acknowledgements{I thank Nicola Prezza for pointing out the paper \cite{kim2023}.}

\nolinenumbers 

\usepackage{tikz}
\usetikzlibrary{arrows,automata,positioning}

\DeclareMathOperator*{\argmin}{arg\,min}

\hideLIPIcs

\begin{document}

\maketitle

\begin{abstract}
In the past thirty years, numerous algorithms for building the suffix array of a string have been proposed. In 2021, the notion of suffix array was extended from strings to DFAs, and it was shown that the resulting data structure can be built in $ O(m^2 + n^{5/2}) $ time, where $ n $ is the number of states and $ m $ is the number of edges [SODA 2021]. Recently, algorithms running in $ O(mn) $ and $ O(n^2\log n) $ time have been described [CPM 2023].

In this paper, we improve the previous bounds by proposing an $ O(n^2) $ recursive algorithm inspired by Farach's algorithm for building a suffix tree [FOCS 1997]. To this end, we provide insight into the rich lexicographic and combinatorial structure of a graph, so contributing to the fascinating journey which might lead to solve the long-standing open problem of building the suffix tree of a graph.

\end{abstract}

\section{Introduction}

The \emph{suffix tree} \cite{weiner1973} of a string is a versatile data structure introduced by Weiner in 1973 which allows solving a myriad of combinatorial problems, such as determining whether a pattern occurs in the string, computing matching statistics, searching for regular expressions, computing the Lempel-Ziv decomposition of the string and finding palindromes. The book by Gusfield \cite{gusfield1997} devotes almost 150 pages to the applications of suffix trees, stressing the importance of these applications in bioinformatics. However, the massive increase of genomic data in the last decades requires space-efficient data structures able to efficiently support pattern matching queries, and the space consumption of suffix trees is too high. In 1990, Manber and Myers invented    \emph{suffix arrays} \cite{manber1993} as a space-efficient alternative to suffix trees. While suffix arrays do not have the full functionality of suffix trees, they still allow solving pattern matching queries. Suffix arrays started a new chapter in data compression, which culminated in the invention of data structures closely related to suffix arrays, notably, the Burrows-Wheeler Transform \cite{burrows1994} and the FM-index \cite{ferragina2000, ferraginajacm2005}, which have heavily influenced sequence assembly \cite{simpson2010}.

The impact of suffix arrays has led to a big effort in the attempt of designing efficient algorithms to construct suffix arrays, where ``efficient'' refers to various metrics (worst-case running time, average running time, space, performance on real data and so on); see \cite{puglisi2007} for a comprehensive survey on the topic. Let us focus on worst-case running time. Manber and Myers build the suffix array of a string of length $ n $ in $ O(n \log n) $ by means of a \emph{prefix-doubling} algorithm \cite{manber1993}. In 1997, Farach proposed a recursive algorithm to build the suffix \emph{tree} of a string in linear time for integer alphabets \cite{farach1997}. In the following years, the recursive paradigm of Farach's algorithm was used to developed a multitude of linear-time algorithms for building the suffix array \cite{ko2005, karkkainen2006, kim2005, na2005}. All these algorithms carefully exploit the lexicographic struture of the suffixes of a string, recursively reducing the problem of computing the suffix array of a string to the problem of computing the suffix array of a smaller string (\emph{induced sorting}).

The problem of solving pattern matching queries not only on strings, but also on labeled graphs, is an active topic of research. Recently, Equi et al. showed that no algorithm can solve pattern matching queries \emph{on arbitrary graphs} in $ O(m^{1 - \epsilon} |P|) $ time or $ O(m |P|^{1 - \epsilon})$ (where $ m $ is the number of edges, $ P $ is the pattern and $ \epsilon > 0 $), unless the Orthogonal Vectors hypothesis fails \cite{equi2023, equi2021}. On the other hand, over the years the idea of (lexicographically) sorting the suffixes of a string has been generalized to graphs, thus leading to compact data structures that are able to support pattern matching queries on graphs. The mechanism behind the suffix array, the Burrows-Wheeler Transform and the FM-index was first generalized to trees \cite{ferragina2005, ferraginajacm2009}; later on, it was generalized to De Brujin graphs \cite{BOSS, GCSA} (which can be used for Eulerian sequence assembly \cite{idury1995}). Subsequently, it was extended to the so-called Wheeler graphs \cite{gagie2017, alanko2020}, and finally to arbitrary graphs and automata \cite{cotumaccio2021, cotumaccio2022}. The idea of lexicographically sorting the strings reaching the states of an automaton has also deep theoretical consequences in automata theory: for example, it leads to a parametrization of the powerset construction, which implies fixed-parameter tractable algorithms for PSPACE-complete problems such as deciding the equivalence of two non-deterministic finite automata (NFAs) \cite{cotumacciojacmpreprint}.

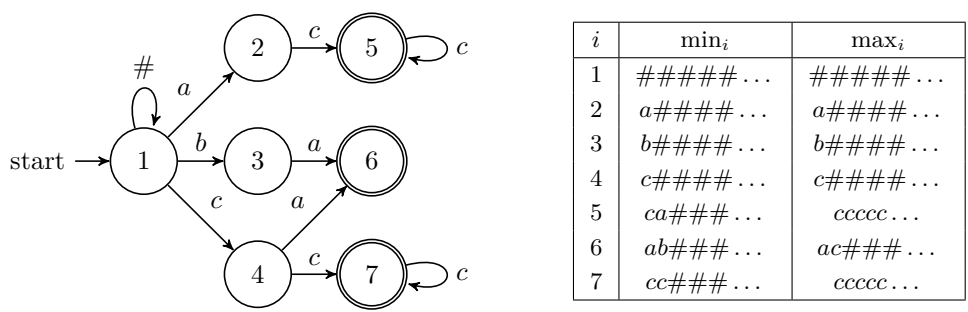
\begin{figure}
     \centering
     \begin{subfigure}[]{0.49\textwidth}
        \centering
        \begin{tikzpicture}[->,>=stealth', semithick, auto, scale=1]
\node[state, initial] (1)    at (0,0)	{$ 1 $};
\node[state] (2)    at (1.5,1.5)	{$ 2 $};
\node[state] (3)    at (1.5,0)	{$ 3 $};
\node[state] (4)    at (1.5,-1.5)	{$ 4 $};
\node[state, accepting] (5)    at (3,1.5)	{$ 5 $};
\node[state, accepting] (6)    at (3,0)	{$ 6 $};
\node[state, accepting] (7)    at (3,-1.5)	{$ 7 $};
\draw (1) edge [] node [] {$ a $} (2);
\draw (1) edge [] node [] {$ b $} (3);
\draw (1) edge [] node [] {$ c $} (4);
\draw (2) edge [] node [] {$ c $} (5);
\draw (3) edge [] node [] {$ a $} (6);
\draw (4) edge [] node [] {$ c $} (7);
\draw (4) edge [] node [] {$ a $} (6);
\draw (5) edge [loop right] node [] {$ c $} (5);
\draw (7) edge [loop right] node [] {$ c $} (7);
\draw (1) edge [loop above] node [] {$ \# $} (1);
\end{tikzpicture}
     \end{subfigure}
     \begin{subfigure}[]{0.49\textwidth}
        \centering
\begin{tabular}{ |c|c|c| } 
 \hline
 $ i $ & $ \min_i $ & $ \max_i $ \\
 \hline
 $ 1 $ & $ \#\#\#\#\#\dots $ & $ \#\#\#\#\#\dots $ \\
$ 2 $ & $ a\#\#\#\#\dots $ & $ a\#\#\#\#\dots $ \\ 
$ 3 $ & $ b\#\#\#\#\dots $ & $ b\#\#\#\#\dots $ \\ 
$ 4 $ & $ c\#\#\#\#\dots $ & $ c\#\#\#\#\dots $ \\ 
$ 5 $ & $ ca\#\#\#\dots $ & $ ccccc\dots $ \\
$ 6 $ & $ ab\#\#\#\dots $ & $ ac\#\#\#\dots $ \\
$ 7 $ & $ cc\#\#\#\dots $ & $ ccccc\dots $ \\ 
 \hline
\end{tabular}
     \end{subfigure}
	\caption{A DFA $ \mathcal{A} $, with the minimum and maximum string reaching each state (we assume $ \# \prec a \prec b \prec c $). The min/max partition is given by $ \{(1, \min), (1, \max) \} < \{(2, \min), (2, \max)\} < \{(6, \min) \} < \{(6, \max) \} < \{(3, \min), (3, \max) \} < \{(4, \min), (4, \max) \} < \{(5, \min) \} < \{(7, \min) \} < \{(5, \max), (7, \max) \} $, meaning that $ \min_1 = \max_1 \prec \min_2 = \max_2  \prec \min_6 \prec \max_6 \prec \min_3 = \max_3 \prec \min_4 = \max_4 \prec \min_5 \prec \min_7 \prec \max_5 = \max_7 $.}
 \label{fig:exampleminmax}
\end{figure}

The case of deterministic finite automata (DFAs) is of particular interest, because in this case the notion of ``suffix array'' of a DFA has a simple interpretation in terms of strings. Assume that there is a fixed total order $ \preceq $ on the alphabet $ \Sigma $ of a DFA $ \mathcal{A} $, and extend $ \preceq $ lexicographically to the set of all infinite strings on $ \Sigma $. Assume that each state has at least one incoming edge. If $ u $ is a state, consider the set $ I_u $ of all infinite strings that can be read starting from $ u $ and following edges \emph{in a backward fashion}, and let $ \min_u $ and $ \max_u $ be the lexicograpically smallest (largest, respectively) string in $ I_u $ (see Figure \ref{fig:exampleminmax}). Consider the partial order $ \preceq_\mathcal{A} $ on the set of all states $ Q $ such that for every $ u, v \in Q $, with $ u \not = v $, it holds $ u \prec_\mathcal{A} v $ if and only if $ \max_u \preceq \min_v $. Then, the partial order $ \preceq_\mathcal{A} $ induces a (partial) permutation of the set of all states that plays the same role of the permutation of text positions induced by the suffix array of a string \cite{cotumaccio2021, kim2023}, so $ \preceq_\mathcal{A} $ is essentially the ``suffix array'' of the DFA. If we are able to compute the partial order $ \preceq_\mathcal{A} $ (and a minimum-size partition of $ Q $ into sets such that the restriction of $ \preceq_\mathcal{A} $ to each set is a \emph{total} order), then we can efficiently and compactly solve pattern matching queries on the DFA by means of techniques that extend the Burrows-Wheeler transform and the FM-index from strings to graphs. As a consequence, we now have to solve the problem of efficiently building the ``suffix array'' of a DFA, that is, the problem of computing $ \preceq_\mathcal{A} $.

The first paper on the topic \cite{cotumaccio2021} presents an algorithm that builds $ \preceq_\mathcal{A} $ in $ O(m^2 + n^{5 / 2}) $ time, where $ n $ is the number of states and $ m $ is the number of edges. In a recent paper \cite{kim2023}, Kim et al. describe two algorithms running in $ O(mn) $ and $ O(n^2 \log n) $ time. The key observation is that, if we build the $ \emph{min/max-partition} $ of the set of all states, then we can determine the partial order $ \preceq_\mathcal{A} $ in linear time by means of a reduction to the interval partitioning problem. Determining the min/max-partition of the set of all states means picking the string $ \min_u $ and $ \max_u $ for every state $ u \in Q $, and sorting these $ 2|Q|$ strings (note that some of these strings may be equal, see Figure \ref{fig:exampleminmax} for an example). As a consequence, we are only left with the problem of determining the  min/max-partition efficiently.

The $ O(n^2 \log n) $ algorithm builds the min/max-partition by generalizing Manber and Myers's $ O(n \log n) $ algorithm from strings to DFAs. However, since it is possible to build the suffix array of a string in $ O(n) $ time, it is natural to wonder whether it is possible to determine the min/max-partition in $ O(n^2) $ time.

In this paper, we show that, indeed, it is possible to build the min/max-partition in $ O(n^2) $ time by adopting a recursive approach inspired by one of the linear-time algorithms that we have mentioned earlier, namely, Ko and Aluru's algorithm \cite{ko2005}. As a consequence, our algorithm is asymptotically faster than all previous algorithms.

A long-standing open problem is whether it is possible to define a suffix tree of a graph. Some recent work \cite{conte2023} suggests that it is possible to define data structures that \emph{simulate} the behavior of a suffix tree by carefully studying the lexicographic structure of the graph (the results in \cite{conte2023} only hold for Wheeler graphs, but we believe that they can be extended to arbitrary graphs). More specifically, it is reasonable to believe that it is possible to generalize the notion of \emph{compressed suffix tree} of a string \cite{sadakane2007} to graphs. A compressed suffix tree is a compressed representation of a suffix tree which consists of some components, including a suffix array. We have already seen that $ \preceq_\mathcal{A} $ generalizes the suffix array to a graph structure, and \cite{conte2023} suggests that the remaining components may also be generalized to graphs. The complements of the suffix tree of a string heavily exploit the lexicographic and combinatorial structure of a string. Since the algorithm that we present in this paper deeply relies on the richness of the lexicographic structure (which becomes even more challenging and surprising when switching from a string setting to a graph setting), we believe that our results also provide a solid conceptual contribution towards extending suffix tree functionality to graphs.

We remark that, a few days after we submitted this paper to arXiv, a new arXiv preprint showed how to determine the min/max partition in $ O(m \log n) $ time, where $ n $ is the number of states and $ m $ is the number of edges (this new arXiv preprint was also accepted for publication \cite{beckeresa}). If the graph underlying the DFA is sparse, then the algorithm in \cite{beckeresa} improves our $ O(n^2) $ algorithm. Since the $ O(m \log n) $ algorithm uses different techniques (it is obtained by adapting Paige and Tarjan's partition refinement algorithm \cite{paige1987}), we are left with the intriguing open problem of determining whether, by possibly combining the ideas behind our algorithm and the algorithm in \cite{beckeresa}, it is possible to build the min/max partition in $ O(m) $ time.

Due to space constraints, all proofs can be found in the appendix.

\section{Preliminaries}

\subsection{Relation with Previous Work}

In the setting of the previous works on the topic \cite{alanko2020, cotumaccio2021, kim2023}, the problem that we want to solve is defined as follows. Consider a deterministic finite automaton (DFA) such that (i) all edges entering the same state have the same label, (ii) each state is reachable from the initial state, (iii) each state is co-reachable, that is, it is either final or it allows reaching a final state, (iv) the initial state has no incoming edges. Then, determine the min/max partition of the set of states (see Section \ref{sec:notation} for the formal definition of min/max partition, and see Figure \ref{fig:exampleminmax} for an example). 
\begin{itemize}
    \item Assumptions (ii) and (iii) are standard assumptions in automata theory, because all states that do not satisfy these assumptions can be removed without changing the accepted language.
    \item In this setting, all the non-initial states have an incoming edge, but the initial state has no incoming edges. This implies for some state $ u $ it may hold $ I_u = \emptyset $ (remember that $ I_u $ is the set of all infinite strings that can be read starting from $ u $ and following edges in a backward fashion, see the introduction), so Kim et al. \cite{kim2023} need to perform a tedious case analysis which also takes finite strings into account in order to define the min/max-partition  (in particular, the minimum and maximum strings reaching the initial state are both equal to the empty string). However, we can easily avoid this complication by means of the same trick used in \cite{conte2023}; we can add a self-loop to the initial state, and the label of the self-loop is a special character $ \# $ \emph{smaller} than any character in the alphabet. Intuitively, $ \# $ plays the same role as the termination character in the Burrows-Wheeler transform of a string, and since $ \# $ is the smallest character, adding this self-loop does not affect the min/max-partition (see \cite{kim2023} for details).
    \item Notice that the initial state and the set of all final states play no role in the definition of the min/max partition; this explains why, more generally, it will be expedient to consider deterministic graphs rather than DFAs (otherwise we would need to artificially add an initial state and add a set of final states when we recursively build a graph in our algorithm). Equivalently, one may assume to work with semiautomata in which the transition function is not necessarily total. This justifies the assumptions that we will make in Section \ref{sec:notation}.
    \item Some recent papers \cite{cotumaccio2022, cotumacciojacmpreprint} have shown that assumptions (i) and (iv) can be removed. The partial order $ \preceq_\mathcal{A} $ is defined analogously, and all the algorithms for building $ \preceq_\mathcal{A} $ that we have mentioned still work. Indeed, if a state $ u $ is reached by edges with the distinct labels, we need to only consider all edges with the smallest label when computing $ \min_u $ and all edges with the largest label when computing $ \max_u $; moreover, we once again assume that the initial state has a self loop labeled $ \# $. The only difference is that assumption (i) implies that $ m \le n^2 $ ($ n $ being the number of states, $ m $ being the number of edges) because each state can have at most $ n $ incoming edges, but this is no longer true if we remove assumption (i). As a consequence, the running time of our algorithm is no longer $ O(n^2) $ but $ O(m + n^2) $ (and the running time of the $ O(n^2 \log n) $ algorithm in \cite{kim2023} becomes $ O(m + n^2 \log n) $) because we still need to process all edges in the DFA. 
\end{itemize}
To sum up, all the algorithms for computing $ \preceq_\mathcal{A} $ work on arbitrary DFAs.

\subsection{Notation and First Definitions}\label{sec:notation}

Let $ \Sigma $ be a finite alphabet. We consider finite, edge-labeled graphs $ G = (V, E) $, where $ V $ is the set of all nodes and $ E \subseteq V \times V \times \Sigma $ is the set of all edges. Up to taking a subset of $ \Sigma $, we assume that all $ c \in \Sigma $ label some edge in the graph. We assume that all nodes have at least one incoming edge, and all edges entering the same node $ u $ have the same label $ \lambda (u) $ (\emph{input consistency}). This implies that an edge $ (u, v, a) \in E $ can be simply denoted as $ (u, v) $, because it must be $ a = \lambda (v) $. In particular, it must be $ |E| \le |V|^2 $ (and so an $ O(|E|) $ algorithm is also a $ O(|V|^2) $ algorithm). If we do not know the $ \lambda (u) $'s, we can easily compute them by scanning all edges. In addition, we always assume that $ G $ is \emph{deterministic}, that is, for every $ u \in V $ and for every $ a \in \Sigma $ there exists at most one $ v \in V $ such that $ (u, v) \in E $ and $ \lambda (v) = a $.

Let $ \Sigma^* $ be the set of all finite strings on $ \Sigma $, and let $ \Sigma^\omega $ be the set of all (countably) right-infinite strings on $ \Sigma $. If $ \alpha \in \Sigma^* \cup \Sigma^\omega $ and $ i \ge 1 $, we denote by $ \alpha[i] \in \Sigma $ the $ i^{\text{th}} $ character of $ \alpha $ (that is, $ \alpha = \alpha [1] \alpha [2] \alpha[3] \dots $). If $ 1 \le i \le j $, we define $ \alpha[i, j] = \alpha[i]\alpha[i + 1]\dots \alpha[j - 1]\alpha[j] $, and if $ j < i $, then $ \alpha[i, j] $ is the empty string $ \epsilon $. If $ \alpha \in \Sigma^*$, then $ |\alpha| $ is length of $ \alpha $; for every $ 0 \le j \le |\alpha| $ the string $ \alpha[1, j] $ is a \emph{prefix} of $ \alpha $, and if $ 0 \le j < |\alpha| $ it is a \emph{strict prefix} of $ \alpha $; analogously, one defines suffixes and strict suffixes of $ \alpha $. An \emph{occurrence} of $ \alpha \in \Sigma^* $ starting at $ u \in V $ and ending at $ u' \in V $ is a sequence of nodes $ u_1, u_2, \dots, u_{|\alpha| + 1} $ of $ V $ such that (i) $ u_1 = u $, (ii) $ u_{|\alpha| + 1} = u' $, (iii) $ (u_{i + 1}, u_i) \in E $ for every $ 1 \le i \le |\alpha| $ and (iv) $ \lambda (u_i) = \alpha[i] $ for every $ 1 \le i \le |\alpha| $. An \emph{occurrence} of $ \alpha \in \Sigma^\omega $ starting at $ u \in V $ is a sequence of nodes $ (u_i)_{i \ge 1} $ of $ V $ such that (i) $ u_1 = u $, (ii) $ (u_{i + 1}, u_i) \in E $ for every $ i \ge 1 $ and (iii) $ \lambda (u_i) = \alpha[i] $ for every $ i \ge 1 $. Intuitively, a string $ \alpha \in \Sigma^* \cup \Sigma^\omega $ has an occurrence starting at $ u \in V $ if we can read $ \alpha $ on the graph starting from $ u $ and following edges \emph{in a backward fashion}.

In the paper, occurrences of strings in $ \Sigma^\omega $ will play a key role, while occurrences of strings in $ \Sigma^* $ will be used as a technical tool. For every $ u \in V $, we denote by $ I_u $ the set of all strings in $ \Sigma^\omega $ admitting an occurrence starting at $ u $. Since every node has at least one incoming edge, then $ I_u \not = \emptyset $.

A \emph{total order} $ \le $ on a set $ V $ if a reflexive, antisymmetric and transitive relation on $ V $. If $ u, v \in V $, we write $ u < v $ if $ u \le v $ and $ u \not = v $.

Let $ \preceq $ be a fixed total order on $ \Sigma $. We extend $ \preceq $ to $ \Sigma^* \cup \Sigma^\omega $ \emph{lexicographically}. It is easy to show that in every $ I_u $ there is a lexicographically smallest string $ \min_u $ and a lexicographically largest string $ \max_u $ (for example, it follows from \cite[Observation 8]{kim2023}).

We will often use the following immediate observation. Let $ u \in V $, and let $ (u_i)_{i \ge 1 } $ be an occurrence of $ \min_u $. Fix $ i \ge 1 $. Then, $ (u_j)_{j \ge i} $ is an occurrence of $ \min_{u_i} $, and $ \min_u = \min_u[1, i - 1] \min_{u_i} $.

Let $ V' \subseteq V $. Let $ \mathcal{A} $ be the unique partition of $ V' $ and let $ \le $ be the unique total order on $ \mathcal{A} $ such that, for every $ I, J \in \mathcal{A} $ and for every $ u \in I $ and $ v \in J $, (i) if $ I = J $, then $ \min_u = \min_v $ and (ii) if $ I < J $, then $ \min_u \prec \min_v $. Then, we say that $ (\mathcal{A}, \le) $, or more simply $ \mathcal{A} $, is the $ \emph{min-partition} $ of $ V' $. The $ \emph{max-partition} $ of $ V' $ is defined analogously. Now, consider the set $ V' \times \{\min, \max \} $, and define $ \rho ((u, \min)) = \min_u $ and $ \rho ((u, \max)) = \max_u $ for every $ u \in V' $. Let $ \mathcal{B} $ be the unique partition of $ V' \times \{\min, \max \} $ and let $ \le $ be the unique total order on $ \mathcal{B} $ such that, for every $ I, J \in \mathcal{B} $ and for every $ x \in I $ and $ y \in J $, (i) if $ I = J $, then $ \rho(x) = \rho(y) $ and (ii) if $ I < J $, then $ \rho (x) \prec \rho(y) $. Then, we say that $ (\mathcal{B}, \le) $, or more simply $ \mathcal{B} $, is the $ \emph{min/max-partition} $ of $ V' $.

The main result of this paper will be proving that the min/max partition of $ V $ can be determined in $ O(|V|^2) $ time.

\subsection{Our Approach}

Let $ G = (V, E) $ be a graph. We will first show how to build the min-partition of $ V $ in $ O(n^2) $ time, where $ n = |V| $ (Section \ref{sec:computingmin}); then, we will show how the algorithm can be adapted so that it builds the min/max-partition in $ O(n^2) $ time (Section \ref{sec:minmax}).

In order to build a min-partition of $ V $, we will first classify all minima into three categories (Section \ref{sec:classifyingstrings}), so that we can split $ V $ into three pairwise-disjoint sets $ V_1, V_2, V_3 $. Then, we will show that in $ O(n^2) $ time:
\begin{itemize}
    \item we can compute $ V_1, V_2, V_3 $ (Section \ref{sec:minimaclassification});
    \item we can define a graph $ \bar{G} = (\bar{V}, \bar{E}) $ having $ |V_3| $ nodes  (Section \ref{sec:recursivestep});
    \item assuming that we have already determined the min-partition of $ \bar{V} $, we can determine the min-partition of $ V $ (Section \ref{sec:merging}).
\end{itemize}
Analogously, in $ O(n^2) $ time we can reduce the problem of determining the min-partition of $ V $ to the problem of determining the min-partition of the set of all nodes of a graph having $ |V_1| $ (not $ |V_3| $) nodes (Section \ref{sec:complementary}). As a consequence, since $ \min \{|V_1|, |V_3|\} \le |V| / 2 = n / 2 $, we obtain a recursive algorithm whose running time is given by the recurrence:
\begin{equation*}
    T(n) = T(n / 2) + O(n^2)
\end{equation*}
and we conclude that the running time of our algorithm is $ O(n^2) $.

\section{Classifying Strings}\label{sec:classifyingstrings}

In \cite{ko2005}, Ko and Aluru divide the suffixes of a string into two groups. Here we follow an approach purely based on stringology, without fixing a string or a graph from the start. We divide the strings of $ \Sigma^\omega $ into three groups, which we call group 1, group 2 and group 3 (Corollary \ref{cor:typesproperties} provides the intuition behind this choice).

\begin{definition}
    Let $ \alpha \in \Sigma^\omega $. Let $ a \in \Sigma $ and $ \alpha' \in \Sigma^\omega $ such that $ \alpha = a \alpha' $. Then, we define $ \tau (\alpha) $ as follows:
    \begin{enumerate}
        \item $ \tau(\alpha) = 1 $ if $ \alpha' \prec \alpha $.
        \item $ \tau(\alpha) = 2 $ if $ \alpha' = \alpha $.
        \item $ \tau(\alpha) = 3 $ if $ \alpha \prec \alpha' $. 
    \end{enumerate}
\end{definition}

We will constantly use the following characterization.

\begin{lemma}\label{lem:typeequivalent}
    Let $ \alpha \in \Sigma^\omega $. Let $ a \in \Sigma $ and $ \alpha' \in \Sigma^\omega $ such that $ \alpha = a \alpha' $. Then:
    \begin{enumerate}
        \item $ \tau(\alpha) = 2 $ if and only if $ \alpha' = a^\omega $, if and only if $ \alpha = a^\omega $.
        \item $ \tau(\alpha) \not = 2 $ if and only if $ \alpha' \not = a^\omega $, if and only if $ \alpha \not = a^\omega $.
    \end{enumerate}
    Assume that $ \tau(\alpha) \not = 2 $. Then, there exist unique $ c \in \Sigma \setminus \{a \} $, $ \alpha'' \in \Sigma^\omega $ and $ i \ge 0 $ such that $ \alpha' = a^i c \alpha'' $ (and so $ \alpha = a^{i + 1} c \alpha'' $). Moreover:
    \begin{enumerate}
        \item $ \tau(\alpha) = 1 $ if and only if $ c \prec a $, if and only if $ \alpha' \prec a^\omega $, if and only if $ \alpha \prec a^\omega $.
        \item $ \tau(\alpha) = 3 $ if and only if $ a \prec c $, if and only if $ a^\omega \prec \alpha' $, if and only if $ a^\omega \prec \alpha $,
    \end{enumerate}
\end{lemma}

The following corollary will be a key ingredient in our recursive approach.

\begin{corollary}\label{cor:typesproperties}
    Let $ \alpha, \beta \in \Sigma^\omega $. Let $ a, b \in \Sigma $ and $ \alpha', \beta' \in \Sigma^\omega $ such that $ \alpha = a \alpha' $ and $ \beta = b \beta'$. Then:
    \begin{enumerate}
        \item If $ a = b $ and $ \tau (\alpha) = \tau (\beta) = 2 $, then $ \alpha = \beta $.
        \item If $ a = b $ and $ \tau (\alpha) < \tau (\beta) $, then $ \alpha \prec \beta $. Equivalently, if $ a = b $ and $ \alpha \preceq \beta $, then $ \tau (\alpha) \le \tau (\beta) $.
    \end{enumerate}
\end{corollary}

\section{Computing the min-partition}\label{sec:computingmin}

\begin{figure}
     \centering
     \begin{subfigure}[]{0.49\textwidth}
        \centering
        \begin{tikzpicture}[->,>=stealth', semithick, auto, scale=1]
\node[state] (1)    at (0,0)	{$ 1 $};
\node[state] (2)    at (1.5,1.5)	{$ 2 $};
\node[state] (3)    at (1.5,0)	{$ 3 $};
\node[state] (4)    at (1.5,-1.5)	{$ 4 $};
\node[state] (5)    at (3,1.5)	{$ 5 $};
\node[state] (6)    at (3,0)	{$ 6 $};
\node[state] (7)    at (3,-1.5)	{$ 7 $};
\draw (1) edge [] node [] {$ a $} (2);
\draw (1) edge [] node [] {$ b $} (3);
\draw (1) edge [] node [] {$ c $} (4);
\draw (2) edge [] node [] {$ c $} (5);
\draw (3) edge [] node [] {$ a $} (6);
\draw (4) edge [] node [] {$ c $} (7);
\draw (4) edge [] node [] {$ a $} (6);
\draw (5) edge [loop right] node [] {$ c $} (5);
\draw (7) edge [loop right] node [] {$ c $} (7);
\draw (1) edge [loop above] node [] {$ \# $} (1);
\end{tikzpicture}
     \end{subfigure}
     \begin{subfigure}[]{0.49\textwidth}
        \centering
\begin{tabular}{ |c|c|c| } 
 \hline
 $ i $ & $ \min_i $ & $ \tau(i) \; (= \tau(\min_i)) $ \\
 \hline
 $ 1 $ & $ \#\#\#\#\#\dots $ & $ 2 $  \\
$ 2 $ & $ a\#\#\#\#\dots $ & $ 1 $ \\ 
$ 3 $ & $ b\#\#\#\#\dots $ & $ 1 $ \\ 
$ 4 $ & $ c\#\#\#\#\dots $ & $ 1 $ \\ 
$ 5 $ & $ ca\#\#\#\dots $ & $ 1 $ \\
$ 6 $ & $ ab\#\#\#\dots $ & $ 3 $ \\
$ 7 $ & $ cc\#\#\#\dots $ & $ 1 $ \\ 
 \hline
\end{tabular}
     \end{subfigure}
	\caption{The graph from Figure \ref{fig:exampleminmax}, with the values $ \min_i $'s and $ \tau(i) $'s.}
 \label{fig:types}
\end{figure}
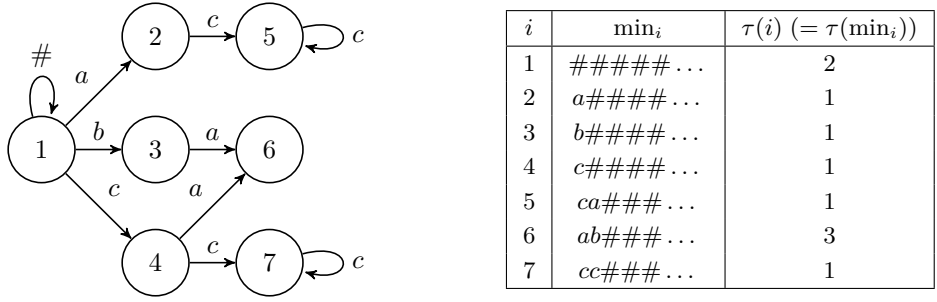

Let $ G = (V, E) $ be a graph. We will prove that we can compute the min-partition of $ V $ in $ O(|V|^2) $ time. In this section, for every $ u \in V $ we define $ \tau (u) = \tau (\min_u) $ (see Figure \ref{fig:types}).

Let $ u \in V $, and let $ (u_i)_{i \ge 1} $ be an occurrence of $ \min_u $ starting at $ u $. It is immediate to realize that (i) if $ \tau (u) = 1 $, then $ \lambda (u_2) \preceq \lambda (u_1) $, (ii) if $ \tau (u) = 2 $, then $ \lambda (u_k) = \lambda (u_1) $ for every $ k \ge 1 $ and (iii) if $ \tau (u) = 3 $, then $ \lambda (u_1) \preceq \lambda (u_2) $.

As a first step, let us prove that without loss of generality we can remove some edges from $ G $ without affecting the min/max-partition. This preprocessing will be helpful in Lemma \ref{lem:runningtimderivedgraph}.

\begin{definition}
    Let $ G = (V, E) $ be a graph. We say that $ G $ is \emph{trimmed} if it contains no edge $ (u, v) \in E $ such that $ \tau (v) = 1 $ and $ \lambda (v) \prec \lambda (u) $.
\end{definition}

In order to simplify the readability of our proofs, we will not directly remove some edges from $ G = (V, E) $, but we will first build a copy of $ G $ where every node $ u $ is a mapped to a node $ u^* $, and then we will trim the graph. In this way, when we write $ \min_u $ and $ \min_{u^*} $ it will be always clear whether we refer to the original graph or the trimmed graph. We will use the same convention in Section \ref{sec:recursivestep} when we define the graph $ \bar{G} = (\bar{V}, \bar{E}) $ that we will use for the recursive step.

\begin{lemma}\label{lem:trimming}
    Let $ G = (V, E) $ be a graph. Then, in $ O(|E|) $ time we can build a trimmed graph $ G^* =(V^*, E^*) $, with $ V^* = \{u^* \;|\; u \in V \} $, such that for every $ u \in V $ it holds $ \min_{u^*} = \min_{u} $. In particular, $ \tau(u^*) = \tau (u) $ for every $ u \in V $.
\end{lemma}

\subsection{Classifying Minima}\label{sec:minimaclassification}

Let us first show how to compute all $ u \in V $ such that $ \tau (u) = 1 $.

\begin{lemma}\label{lem:nextL}
    Let $ G = (V, E) $ be a graph, and let $ u, v \in V $.
    \begin{enumerate}
        \item If $ (u, v) \in E $ and $ \lambda (u) \prec \lambda (v) $, then $ \tau (v) = 1 $.
        \item If $ (u, v) \in E $, $ \lambda (u) = \lambda (v) $ and $ \tau(u) = 1 $, then $ \tau(v) = 1 $.
    \end{enumerate}
\end{lemma}

\begin{corollary}\label{cor:findingt(u)=1}
    Let $ G = (V, E) $ be a graph, and let $ u \in V $. Then, $ \tau (u) = 1 $ if and only if there exist $ k \ge 2 $ and $ z_1, \dots, z_k \in V $ such that (i) $ (z_i, z_{i + 1}) \in E $ for every $ 1 \le i \le k - 1$, (ii) $ z_k = u $, (iii) $ \lambda (z_1) \prec \lambda (z_2) $ and (iv) $ \lambda (z_2) = \lambda (z_3) = \dots = \lambda (z_k) $.
\end{corollary}

Corollary \ref{cor:findingt(u)=1} yields an algorithm to decide whether $ u \in V $ is such that $ \tau (u) = 1 $.

\begin{corollary}\label{cor:type1}
    Let $ G = (V, E) $ be a graph. We can determine all $ u \in V $ such that $ \tau (u) = 1 $ in time $ O(|E|) $.
\end{corollary}

Now, let us show how to determine all $ u \in V $ such that $ \tau (u) = 2 $. We can assume that we have already determined all $ u \in V $ such that $ \tau (u) = 1 $. 

\begin{lemma}\label{lem:characterizationtype2}
     Let $ G = (V, E) $ be a graph, and let $ u \in V $ such that $ \tau (u) \not = 1 $. Then, we have $ \tau (u) = 2 $ if and only if there exist $ k \ge 2 $ and $ z_1, \dots, z_k \in V $ such that (i) $ (z_{i + 1}, z_i) \in E $ for every $ 1 \le i \le k - 1 $, (ii) $ z_1 = u $, (iii) $ z_k = z_j $ for some $ 1 \le j \le k - 1 $ and (iv) $ \lambda (z_1) = \lambda (z_2) = \dots = \lambda (z_k) $.

     In particular, such $ z_1, \dots, z_k \in V $ must satisfy $ \tau (z_i) = 2 $ for every $ 1 \le i \le k $.
\end{lemma}

\begin{corollary}\label{cor:type2}
     Let $ G = (V, E) $ be a graph. We can determine all $ u \in V $ such that $ \tau (u) = 2 $ in time $ O(|E|) $.   
\end{corollary}

From Corollary \ref{cor:type1} and Corollary \ref{cor:type2} we immediately obtain the following result.

\begin{corollary}\label{cor:computingstates}
    Let $ G = (V, E) $ be a graph. Then, in time $ O(|E|) $ we can compute $ \tau (u) $ for every $ u \in V $.
\end{corollary}

\subsection{Recursive Step}\label{sec:recursivestep}

Let us sketch the general idea to build a smaller graph for the recursive step. We consider each $ u \in V $ such that $ \tau (u) = 3 $, and we follow edges in a backward fashion, aiming to determine a prefix of $ \min_u $. As a consequence, we discard edges through which no occurrence of $ \min_u $ can go, and by Corollary \ref{cor:typesproperties} we can restrict our attention to the nodes $ v $ such that $ \tau (v) $ is minimal. We proceed like this until we encounter nodes $ v' $ such that $ \tau (v') = 3 $.

Let us formalize our intuition. We will first present some properties that the occurrences of a string $ \min_u $ must satisfy.

\begin{lemma}\label{lem:comparingminima}
    Let $ G = (V, E) $ be a graph. Let $ u, v \in V $ be such that $ \min_u = \min_v $. Let $ (u_i)_{i \ge 1} $ be an occurrence of $ \min_u $ and let $ (v_i)_{i \ge 1} $ be an occurrence of $ \min_v $. Then:
    \begin{enumerate}
        \item $ \lambda (u_i) = \lambda (v_i) $ for every $ i \ge 1 $.
        \item $ \min_{u_i} = \min_{v_i} $ for every $ i \ge 1 $.
        \item $ \tau(u_i) = \tau(v_i) $ for every $ i \ge 1 $.
    \end{enumerate}
    In particular, the previous results hold if $ u = v $ and $ (u_i)_{i \ge 1} $ and $ (v_i)_{i \ge 1} $ are two distinct occurrences of $ \min_u $.
\end{lemma}

\begin{lemma}\label{lem:minpairwisedistinct}
Let $ G = (V, E) $ be a graph. Let $ u \in V $ and let $ (u_i)_{i \ge 1} $ an occurrence of $ \min_u $ starting at $ u $. Let $ k \ge 1 $ be such that $ \tau (u_1) = \tau (u_2) = \dots = \tau (u_{k - 1}) = \tau (u_k) \not = 2 $. Then, $ u_1, \dots, u_k $ are pairwise distinct. In particular, $ k \le |V| $.
\end{lemma}

The previous results allow us to give the following definition.

\begin{definition}
    Let $ G = (V, E) $ be a graph. Let $ u \in V $ such that $ \tau (u) = 3 $. Let $ \ell_u $ to be the smallest integer $ k \ge 2 $ such that $ \tau (u_k) \ge 2 $, where $ (u_i)_{i \ge 1} $ is an occurrence of $ \min_u $ starting at $ u $.
\end{definition}

Note that $ \ell_u $ is well-defined, because (i) it cannot hold $ \tau (u_k) = 1 $ for every $ k \ge 2 $ by Lemma \ref{lem:minpairwisedistinct} (indeed, if $ \tau (u_2) = 1 $, then $ (u_i)_{i \ge 2} $ is an occurrence of $ \min_{u_2} $ starting at $ u_2 $, and by Lemma \ref{lem:minpairwisedistinct} there exists $ 2 \le k \le |V| + 2 $ such that $ \tau(u_k) \not = 1 $) and (ii) $ \ell_u $ does not depend on the choice of $ (u_i)_{i \ge 1} $ by Lemma \ref{lem:comparingminima}. In particular, it must be $ \ell_u \le |V| + 1 $ because $ u_1, u_2, \dots, u_{\ell_u - 1} $ are pairwise distinct ($ u_1 $ is distinct from $ u_2, \dots, u_{\ell_u - 1} $ because $ \tau (u_1) = 3 $ and $ \tau (u_2) = \tau (u_3) = \dots \tau (u_{\ell_u - 1}) = 1 $ by the minimality of $ \ell_u $).

\begin{lemma}\label{lem:charatersdecrease}
    Let $ G = (V, E) $ be a graph. Let $ u \in V $ such that $ \tau (u) = 3 $. Then, $ \min_u [i + 1] \preceq \min_u[i] $ for every $ 2 \le i \le \ell_u - 1 $. In particular, if $ 2 \le i \le j \le \ell_u $, then $ \min_u [j] \preceq \min_u [i] $.
\end{lemma}

If $ R \subseteq Q $ is a nonempty set of nodes such that for every $ u, v \in R $ it holds $ \lambda (u) = \lambda (v) $, we define $ \lambda(R) = \lambda (u) = \lambda (v) $. If $ R \subseteq Q $ is a nonempty set of nodes such that for every $ u, v \in R $ it holds $ \tau (u) = \tau (v) $, we define $ \tau(R) = \tau (u) = \tau (v) $.

Let $ R \subseteq Q $ be a nonempty set of states. Let $ \mathtt{F}(R) = \argmin_{u \in R'} \tau(u) $, where $ R' = \argmin_{v \in R} \lambda (v) $. Notice that $ \mathtt{F}(R) $ is nonempty, and both $ \lambda(\mathtt{F}(R)) $ and $ \tau(\mathtt{F}(R)) $ are well-defined. In other words, $ \mathtt{F}(R) $ is obtained by first considering  the subset $ R' \subseteq \mathtt{F}(R) $ of all nodes $ v $ such that $ \lambda (v) $ is as small as possible, and then considering the subset of $ R' $ of all nodes $ v $ such that $ \tau (v) $ is as small as possible. This is consistent with our intuition on how we should be looking for a prefix of $ \min_u $.

Define:
\begin{equation*}
    G_i (u) = 
    \begin{cases}
        \{ u \} & \text{ if $ i = 1 $;} \\
        \mathtt{F}(\{v' \in Q \;|\; (\exists v \in G_{i - 1}(u))((v', v) \in E)\}) \setminus \bigcup_{j = 2}^{i - 1} G_j (u) & \text{ if $ 1 < i \le \ell_u $.} \\
    \end{cases}
\end{equation*}

Notice that we also require that a node in $ G_i (u) $ has not been encountered before. Intuitively, this does not affect our search for a prefix of $ \min_u $ because, if we met the same node twice, then we would have a cycle where all edges are equally labeled (because by Lemma \ref{lem:charatersdecrease} labels can only decrease), and since $ \tau (G_i (u)) = 1 $ for every $ 2 \le i \le \ell_u - 1 $, then no occurrence of the minimum can go through the cycle because if we remove the cycle from the occurrence we obtain a smaller string by Lemma \ref{lem:typeequivalent}.

The following technical lemma is crucial to prove that our intuition is correct.

\begin{lemma}\label{lem:propertiesexploringminimum}
    Let $ G = (V, E ) $ be a graph. Let $ u \in V $ such that $ \tau (u) = 3 $.
    \begin{enumerate}
        \item $ G_i (u) $ is well-defined and nonempty for every $ 1 \le i \le \ell_u $.
        \item Let $ (u_i)_{i \ge 1} $ be an occurrence of $ \min_u $ starting at $ u $. Then, $ u_i \in G_i (u) $ for every $ 1 \le i \le \ell_u $. In particular, $ \tau(u_i) = \tau(G_i(u))$ and $ \min_u[i] = \lambda(u_i) = \lambda(G_i(u))$ for every $ 1 \le i \le \ell_u $.
        \item For every $ 1 \le i \le \ell_u $ and for every $ v \in G_i (u) $ there exists an occurrence of $ \min_u[1, i - 1] $ starting at $ u $ and ending at $ v $.
    \end{enumerate}
\end{lemma}

Let $ u \in V $ such that $ \tau (u) = 3 $. We define:
\begin{itemize}
    \item $ \gamma_u = \min_u[1, \ell_u] $;
    \item $ \mathtt{t}_u = \tau(G_{\ell_u} (u)) \in \{2, 3 \} $
\end{itemize}

Now, in order to define the smaller graph for the recursive step, we also need a new alphabet $ (\Sigma', \preceq') $, which must be defined consistently with the mutual ordering of the minima. The next lemma yields all the information that we need.

\begin{lemma}\label{lem_motivatingorder}
    Let $ G = (V, E) $ be a graph. Let $ u, v \in V $ such that $ \tau(u) = \tau(v) = 3 $. Assume that one of the following statements is true:
    \begin{enumerate}
        \item $ \gamma_u $ is not a prefix of $ \gamma_v $ and $ \gamma_u \prec \gamma_v $.
        \item $ \gamma_u = \gamma_v $, $ \mathtt{t}_u = 2 $ and $ \mathtt{t}_v = 3 $.
        \item $ \gamma_v $ is a strict prefix of $ \gamma_u $.
    \end{enumerate}
    Then, $ \min_u \prec \min_v $.
    
    Equivalently, if $ \min_u \preceq \min_v $, then one the following is true: (i) $ \gamma_u $ is not a prefix of $ \gamma_v $ and $ \gamma_u \prec \gamma_v $; (ii) $ \gamma_u = \gamma_v $ and $ \mathtt{t}_u \le \mathtt{t}_v $; (iii) $ \gamma_v $ is a strict prefix of $ \gamma_u $.
\end{lemma}

Now, let $ \Sigma' = \{(\gamma_u, \mathtt{t}_u) \;|\; u \in V, \tau (u) = 3 \} $, and let $ \preceq' $ be the total order on $ \Sigma' $ such that for every distinct $ (\alpha, x), (\beta, y) \in \Sigma' $, it holds $ (\alpha, x) \prec' (\beta, y) $ if and only if one of the following is true:
\begin{enumerate}
    \item $ \alpha $ is not a prefix of $ \beta $ and $ \alpha \prec \beta $.
    \item $ \alpha = \beta $, $ x = 2 $ and $ y = 3 $.
    \item $ \beta $ is a strict prefix of $ \alpha $.
\end{enumerate}
It is immediate to verify that $ \preceq' $ is a total order: indeed, $ \preceq' $ is obtained (i) by first comparing the $ \gamma_u $'s using the variant of the (total) lexicographic order on $ \Sigma^* $ in which a string is smaller than every strict prefix of it and (ii) if the $ \gamma_u $'s are equal by comparing the $ \mathtt{t}_u $'s, which are elements in $ \{2, 3 \} $.

Starting from $ G = (V, E) $, we define a new graph $ \bar{G} = (\bar{V}, \bar{E}) $ as follows:
\begin{itemize}
    \item $ \bar{V} = \{\bar{u} \; | \; u \in V, \tau(u) = 3 \} $.
    \item The new totally-ordered alphabet is $ (\Sigma', \preceq') $.
    \item For every $ \bar{u} \in \bar{V} $, we define $ \lambda(\bar{u}) = (\gamma_u, \mathtt{t}_u) $.
    \item $ \bar{E} = \{(\bar{v}, \bar{v}) \;|\; \mathtt{t}_v = 2 \} \cup \{(\bar{u}, \bar{v}) \;|\; \mathtt{t}_v = 3, u \in G_{\ell_v}(v) \} $.
\end{itemize}

Note that for every $ \bar{v} \in \bar{V} $ such that $ \mathtt{t}_v = 3 $ and for every $ u \in G_{\ell_v}(v) $ it holds $ \tau (u) = \tau(G_{\ell_v}(v)) = \mathtt{t}_v = 3 $, so $ \bar{u} \in \bar{V} $ and $ (\bar{u}, \bar{v}) \in E $. Moreover, $ \bar{G} = (\bar{V}, \bar{E}) $ satisfies all the assumptions about graphs that we use in this paper: (i) all edges entering the same node have the same label (by definition), (ii) every node has at least one incoming edge (because if $ \bar{v} \in \bar{V} $, then $ G_{\ell_v}(v) \not = \emptyset $ by Lemma \ref{lem:propertiesexploringminimum}) and (iii) $ \bar{G} $ is deterministic (because if $ (\bar{u}, \bar{v}), (\bar{u}, \bar{v}') \in \bar{E} $ and $ \lambda(\bar{v}) = \lambda(\bar{v}') $, then $ \gamma_v = \gamma_{v'} $ and $ \mathtt{t}_v = \mathtt{t}_{v'} $, so by the definition of $ \bar{E} $ if $ \mathtt{t}_v = \mathtt{t}_{v'} = 2 $ we immediately obtain $ \bar{v} = \bar{u} = \bar{v}' $, and if $ \mathtt{t}_v = \mathtt{t}_{v}' = 3 $ we obtain $ u \in G_{\ell_v}(v) \cap G_{\ell_{v'}}(v') $; since by Lemma \ref{lem:propertiesexploringminimum} there exist two occurrences of $ \min_v[1, \ell_v - 1] = \gamma_v[1, \ell_v - 1] = \gamma_{v'}[1, \ell_{v'} - 1] = \min_{v'}[1, \ell_{v'} - 1] $ starting at $ v $ and $ v' $ and both ending at $ u $, the determinism of $ G $ implies $ v = v' $ and so $ \bar{v} = \bar{v}' $).

Notice that if $ \bar{v} \in \bar{V} $ is such that $ \mathtt{t}_v = 2 $, then $ I_{\bar{v}} $ contains exactly one string, namely, $ \lambda (\bar{v})^\omega $; in particular, $ \min_{\bar{v}} = \max_{\bar{v}} = \lambda (\bar{v})^\omega $.

When we implement $ G = (V, E) $ and $ \bar{G} = (\bar{V}, \bar{E}) $, we use integer alphabets $ \Sigma = \{0, 1, \dots, |\Sigma| - 1 \} $ and $ \Sigma' = \{0, 1, \dots, |\Sigma'| - 1 \} $; in particular, we will not store $ \Sigma' $ by means of pairs $ (\gamma_u, \mathtt{t}_u) $'s, but we will remap $ \Sigma' $ to an integer alphabet consistently with the total order $ \preceq' $ on $ \Sigma' $, so that the mutual order of the $ \min_{\bar{u}} $'s is not affected.

Let us prove that we can use $ \bar{G} = (\bar{V}, \bar{E}) $ for the recursive step. We will start with some preliminary results.

\begin{lemma}\label{lem:type2equalityminima}
    Let $ G = (V, E) $ be a graph. Let $ u, v \in V $ be such that $ \tau(u) = \tau(v) = 3 $, $ \gamma_u = \gamma_v $ and $ \mathtt{t}_u = \mathtt{t}_v = 2 $. Then, $ \min_u = \min_v $.
\end{lemma}

\begin{lemma}\label{lem:descrptionminimum}
    Let $ G = (V, E) $ be a graph. Let $ u \in V $, and let $ (u_i)_{i \ge 1} $ be an occurrence of $ \min_u $ starting at $ u $. Then, exactly one of the following holds true:
    \begin{enumerate}
        \item There exists $ i_0 \ge 1 $ such that $ \tau(u_i) \not = 2 $ for every $ 1 \le i < i_0 $ and $ \tau(u_i) = 2 $ for every $ i \ge i_0 $.
        \item $ \tau(u_i) \not = 2 $ for every $ i \ge 1 $, and both $ \tau(u_i) = 1 $ and $ \tau(u_i) = 3 $ are true for infinitely many $ i $'s.
    \end{enumerate}
\end{lemma}

Crucially, the next lemma establishes a correspondence between minima of nodes in $ G = (V, E) $ and minima of nodes in $ \bar{G} = (\bar{V}, \bar{E}) $.

\begin{lemma}\label{ref:minimumquotientgraph}
Let $ G = (V, E) $ be a graph. Let $ u \in V $ such that $ \tau(u) = 3 $. Let $ (u_i)_{i \ge 1} $ be an occurrence of $ \min_u $ starting at $ u $. Let $ (u'_i)_{i \ge 1} $ be the infinite sequence of nodes in $ V $ obtained as follows. Consider $ L = \{k \ge 1 \;|\; \tau(u_k) = 3 \} $, and for every $ i \ge 1 $, let $ j_i \ge 1 $ be the $ i^{\text{th}} $ smallest element of $ L $, if it exists. For every $ i \ge 1 $ such that $ j_i $ is defined, let $ u'_i = u_{j_i} $, and if $ i \ge 1 $ is such that $ j_i $ is not defined (so $ L $ is a finite set), let $ u'_i = u'_{|L|} $.
Then, $ (\bar{u}'_i)_{i \ge 1} $ is an occurrence of $ \min_{\bar{u}} $ starting at $ \bar{u} $ in $ \bar{G} = (\bar{V}, \bar{E}) $.
\end{lemma}

The following theorem shows that our reduction to $ \bar{G} = (\bar{V}, \bar{E}) $ is correct.

\begin{theorem}\label{theor:reductiontosmaller}
    Let $ G = (V, E) $ be a graph. Let $ u, v \in V $ be such that $ \tau(u) = \tau(v) = 3 $.
    \begin{enumerate}
        \item If $ \min_u = \min_v $, then $ \min_{\bar{u}} = \min_{\bar{v}} $.
        \item If $ \min_u \prec \min_v $, then $ \min_{\bar{u}} \prec' \min_{\bar{v}} $.
    \end{enumerate}
\end{theorem}

Since $ \preceq $ is a total order (so exactly one among $ \min_u \prec \min_v $, $ \min_u = \min_v $ and $ \min_v \prec \min_u $ holds true), from Theorem \ref{theor:reductiontosmaller} we immediately obtain the following result.

\begin{corollary}
    Let $ G = (V, E) $ be a graph. Let $ u, v \in V $ be such that $ \tau(u) = \tau(v) = 3 $.
    \begin{enumerate}
        \item It holds $ \min_u = \min_v $ if and only if $ \min_{\bar{u}} = \min_{\bar{v}} $.
        \item It holds $ \min_u \prec \min_v $ if and only if $ \min_{\bar{u}} \prec' \min_{\bar{v}} $.
    \end{enumerate}
    In particular, if we have the min-partition of $ \bar{V} $ (with respect to $ \bar{G} $), then we also have the min-partition of $ \{u \in V \;|\; \tau (u) = 3 \} $ (with respect to $ G $).
\end{corollary}

Lastly, we show that our reduction to $ \bar{G} = (\bar{V}, \bar{E}) $ can be computed within $ O(n^2) $ time.
 
\begin{lemma}\label{lem:runningtimderivedgraph}
    Let $ G = (V, E) $ be a trimmed graph. Then, we can build $ \bar{G} = (\bar{V}, \bar{E}) $ in $ O(|V|^2) $ time.
\end{lemma}

\subsection{Merging}\label{sec:merging}

We want to determine the min-partition $ \mathcal{A} $ of $ V $, assuming that we already have the min-partition $ \mathcal{B}$ of $ \{u \in V \;|\; \tau (u) = 3 \} $.

First, note that we can easily build the min-partition $ \mathcal{B}' $ of $ \{u \in V \;|\; \tau (u) = 2 \} $. Indeed, if $ \tau (u) = 2 $, then $ \min_u = \lambda(u)^\omega $ by Lemma \ref{lem:typeequivalent}. As a consequence, if $ \tau(u) = \tau (v) = 2 $, then (i) $ \min_u = \min_v $ if and only if $ \lambda (u) = \lambda (v) $ and (ii) $ \min_u \prec \min_v $ if and only if $ \lambda (u) \prec \lambda (v) $, so we can build  $ \mathcal{B}' $ in $ O(|V|) $ time by using counting sort.

For every $ c \in \Sigma $ and $ t \in \{1, 2, 3 \} $, let $ V_{c, t} = \{v \in V \;|\; \lambda (v) = c, \tau (v) = t \} $. Consider $ u, v \in V $: (i) if $ \lambda (u) \prec \lambda (v) $, then $ \min_u \prec \min_v $ and (ii) if $ \lambda (u) = \lambda (v) $ and $ \tau (u) < \tau (v) $, then $ \min_u \prec \min_v $ by Corollary \ref{cor:typesproperties}. As a consequence, in order to build $ \mathcal{A} $, we only have to build the min-partition $ \mathcal{A}_{c, t} $ of $ V_{c, t} $, for every $ c \in \Sigma $ and every $ t \in \{1, 2, 3 \} $.


A possible way to implement each $ \mathcal{A}_{c, t} $ is by means of an array $ A_{c, t} $ storing the elements of $ V_{c, t} $, where we also use a special character to delimit the border between consecutive elements of $ A_{c, t} $.

It is immediate to build incrementally $ \mathcal{A}_{c, 3} $ for every $ c \in \Sigma $, from its smallest element to its largest element. At the beginning, $ A_{c, 3} $ is empty for every $ c \in \Sigma $. Then, scan the elements $ I $ in $ \mathcal{B} $ from smallest to largest, and add $ I $ to $ \mathcal{A}_{c, 3} $, where $ c = \lambda (u) $ for any $ u \in I $ (the definition of $ c $ does not depend on the choice of $ u $). We scan $ \mathcal{B} $ only once, so this step takes $ O(|V|) $ time. Analogously, we can build $ \mathcal{A}_{c, 2} $ for every $ c \in \Sigma $ by using $ \mathcal{B}' $.

We are only left with showing how to build $ \mathcal{A}_{c, 1} $ for every $ c \in \Sigma $. At the beginning, each $ A_{c, 1} $ is empty, and we will build each $ \mathcal{A}_{c, 1} $ from its smallest element to its largest element. During this step of the algorithm, we will gradually mark the nodes $ u \in  V $ such that $ \tau (u) = 1 $. At the beginning of the step, no such node is marked, and at the end of the step all these nodes will be marked. Let $ \Sigma = \{c_1, c_2, \dots, c_{\sigma} \} $, with $ c_1 \prec c_2 \prec \dots \prec c_\sigma $. Notice that it must be $ V_{c_1, 1} = \emptyset $, because if there existed $ u \in V_{c_1, 1} $, then it would be $ \min_u \prec c_1^\omega $ by Lemma \ref{lem:typeequivalent} and so $ c_1 $ would not be the smallest character in $ \Sigma $. Now, consider $ V_{c_1, 2} $; we have already fully computed $ A_{c_1, 2} $. Process each $ I $ in $ A_{c_1, 2} $ from smallest to largest, and for every $ c_k \in \Sigma $ compute the set $ J_k $ of all non-marked nodes $ v \in V $ such that $ \tau(v) = 1 $, $ \lambda (v) = c_k $, and $ (u, v) \in E $ for some $ u \in I $. Then, if $ J_k \not = \emptyset $ add $ J_k $ to $ A_{c_k, 1} $ and mark the nodes in $ J_k $.
After processing the elements in $ A_{c_1, 2} $, we process the element in $ A_{c_1, 3} $, $ A_{c_2, 1} $, $ A_{c_2, 2} $, $ A_{c_2, 3} $, $ A_{c_3, 1} $ and so on, in this order. Each $ A_{c_i, t} $ is processed from its (current) smallest element to its (current) largest element. We never remove or modify elements in any $ A_{c, t} $, but we only add elements to the $ A_{c, 1} $'s. More precisely, when we process $ I $ in $ A_{c, t} $, for every $ c_k \in \Sigma $ we compute the set $ J_k $ of all non-marked nodes $ v \in V $ such that $ \tau(v) = 1 $, $ \lambda (v) = c_k $, and $ (u, v) \in E $ for some $ u \in I $ and, if $ J_k \not = \emptyset $, then we add $ J_k $ to $ A_{c_k, 1} $ and we mark the nodes in $ J_k $.

The following lemma shows that our approach is correct. Let us give some intuition. A \emph{prefix} of a min-partition $ \mathcal{C} $ is a subset $ \mathcal{C}' $ of $ \mathcal{C} $ such that, if $ I, J \in \mathcal{C} $, $ I < J $ and $ J \in \mathcal{C}' $, then $ I \in \mathcal{C'} $. Notice that every prefix of $ \mathcal{A} $ is obtained by taking the union of $ \mathcal{A}_{c_1, 2} $, $ \mathcal{A}_{c_1, 3} $, $ \mathcal{A}_{c_2, 1} $, $ \mathcal{A}_{c_2, 2} $, $ \mathcal{A}_{c_2, 3} $, $ \mathcal{A}_{c_3, 1} $, $ \dots $ in this order up to some element $ \mathcal{A}_{c, t} $, where possibly we only pick a prefix of the last element $ \mathcal{A}_{c, t} $. Then, we will show that, when we process $ I $ in $ A_{c, t} $, we have already built the prefix of $ \mathcal{A} $ whose largest element is $ I $. This means that, for every $ v \in J_k $ and for any \emph{any} occurrence $ (v_i)_{i \ge 1} $ of $ \min_v $ starting at $ v $, it must hold that $ v_2 $ is in $ I $.

\begin{lemma}\label{lem:detailsmerging}
    Let $ G = (V, E) $ be a graph. If we know the min-partition of $ \{u \in V \;|\; \tau (u) = 3 \} $, then we can build the min-partition of $ V $ in $ O(|E|) $ time. 
\end{lemma}

\subsection{The Complementary Case}\label{sec:complementary}

We have shown that in $ O(n^2) $ time we can reduce the problem of determining the min-partition of $ V $ to the problem of determining the min-partition of the set of all nodes of a graph having $ |\{u \in V \;|\; \tau (u) = 3 \}| $ nodes. Now, we must show that (similarly) in $ O(n^2) $ time we can reduce the problem of determining the min-partition of $ V $ to the problem of determining the min-partition of the set of all nodes of a graph having $ |\{u \in V \;|\; \tau (u) = 1 \}| $ nodes. This time, building the smaller graph $ \bar{G} = (\bar{V}, \bar{E}) $ will be simpler, because when searching for prefixes the minima we will never meet nodes that we have already met (so the the definition of the $ G_i (u) $'s is simpler and we do not need to trim the graph to stay within the $ O(n^2) $ bound). On the other hand, the merging step will be more complex, because the order in which we will process the $ \mathcal{A}_{c, t} $ will be from largest to smallest ($ \mathcal{A}_{c_\sigma, 2} $, $ \mathcal{A}_{c_\sigma, 1} $, $ \mathcal{A}_{c_{\sigma - 1}, 3} $, $ \mathcal{A}_{c_{\sigma - 1}, 2} $, $ \mathcal{A}_{c_{\sigma - 1}, 1} $, $ \mathcal{A}_{c_{\sigma - 2}, 3} $ and so on) so we will need to update some elements of some $ A_{c, t} $'s to include the information about minima that we may infer at a later stage of the algorithm. We provide the details in the appendix.

\section{Computing the min/max-partition}\label{sec:minmax}

Let $ G = (V, E) $ be a graph. We can build the \emph{max-partition} of $ V $ by simply considering the transpose total order $ \preceq^* $ of $ \preceq $ (the one for which $ a \preceq^* b $ if and only if $ b \preceq a $) and building the min-partition. As a consequence, the algorithm to build the max-partition is entirely symmetrical  to the algorithm to build the min-partition.

Let $ G = (V, E) $ be a graph. Let us show how we can build the min/max-partition of $ V $ in $ O(|V|^2) $ time. Assume that we have two graphs $ G_1 = (V_1, E_1) $ and $ G_2 = (V_2, E_2) $ on the same alphabet $ (\Sigma, \preceq) $, with $ V_1 \cap V_2 = \emptyset $ (we allow $ G_1 $ and $ G_2 $ to possibly be the \emph{null graph}, that is, the graph without vertices). Let $ V'_1 \subseteq V_1 $, $ V'_2 \subseteq V_2 $, $ W = V'_1 \cup V'_2 $, and for every $ u \in W $ define $ \rho (u) = \min_u $ if $ u \in V'_1 $, and $ \rho (u) = \max_u $ if $ u \in V'_2 $. Let $ \mathcal{A} $ be the unique partition of $ W $ and let $ \le $ be the unique total order on $ \mathcal{A} $ such that, for every $ I, J \in \mathcal{A} $ and for every $ u \in I $ and $ u \in J $, (i) if $ I = J $, then $ \rho(u) = \rho(u) $ and (ii) if $ I < J $, then $ \rho (u) \prec \rho(u) $. Then, we say that $ (\mathcal{A}, \le) $, or more simply $ \mathcal{A} $, is the \emph{min/max-partition} of $ (V'_1, V'_2) $. We will show that we can compute the min/max partition of $ (V_1, V_2) $ in $ O((|V_1| + |V_2|)^2) $ time. In particular, if $ G_1 = (V_1, E_1) $ and $ G_2 = (V_1, E_2) $ are two (distinct) copies of the same graph $ G = (V, E) $, then we can compute the min/max-partition of $ V $ in $ O(|V|^2) $ time.

We compute $ \tau (\min_u) $ for every $ u \in V_1 $ and we compute $ \tau (\max_u) $ for every $ u \in V_2 $. If the number of values equal to $ 3 $ is smaller than the number of values equal to 1, then (in time $ O(|V_1|^2 + |V_2|^2) = O((|V_1| + |V_2|)^2) $) we build the graphs $ \bar{G}_1 = (\bar{V}_1, \bar{E}_1) $ and $ \bar{G}_2 = (\bar{V}_2, \bar{E}_2) $ as defined before, where $ \bar{V}_1 = \{\bar{u} \;|\; u \in V_1, \tau (\min_u) = 3 \} $ and $ \bar{V}_2 = \{\bar{u} \;|\; u \in V_2, \tau (\max_u) = 3 \} $, otherwise we consider the complementary case (which is symmetrical). When building $ \bar{G}_1 = (\bar{V}_1, \bar{E}_1) $ and $ \bar{G}_2 = (\bar{V}_2, \bar{E}_2) $, we define a \emph{unique} alphabet $ (\Sigma', \preceq') $ obtained by jointly sorting the $ (\gamma_{\min_u}, \mathtt{t}_{\min_u }) $'s and the $ (\gamma_{\max_u}, \mathtt{t}_{\max_u }) $'s, which is possible because Lemma \ref{lem_motivatingorder} also applies to maxima. Note that $ |\bar{V}_1| + |\bar{V}_2| \le (|V_1| + |V_2|) / 2 $.

Assume that we have recursively obtained the min/max-partition of $ (\bar{V}_1, \bar{V}_2) $ with respect to $ \bar{G}_1 $ and $ \bar{G}_2 $. This yields the min/max-partition of $ (\{u \in V_1 \;|\; \tau(\min_u) = 3 \}, \{u \in V_2 \;|\; \tau(\max_u)= 3 \}) $. Then, we can build the min/max-partition of $ (V_1, V_2) $ by jointly applying the merging step, which is possible because both the merging step for minima and the merging step for maxima require to build the $ \mathcal{A}_{c, 1} $'s by processing $ A_{c_1, 2} A_{c_1, 3} $, $ A_{c_2, 1} $, $ A_{c_2, 2} $, $ A_{c_2, 3} $, $ A_{c_3, 1} $ and so on in this order.

Since we obtain the same recursion as before, we conclude that we can compute the min/max partition of $ (V_1, V_2) $ in $ O((|V_1| + |V_2|)^2) $ time.



\newpage

\bibliography{lipics-v2021-sample-article}

\newpage

\appendix

\section{Proofs from Section \ref{sec:classifyingstrings}} 

\noindent{\textbf{Statement of Lemma \ref{lem:typeequivalent}}.}
Let $ \alpha \in \Sigma^\omega $. Let $ a \in \Sigma $ and $ \alpha' \in \Sigma^\omega $ such that $ \alpha = a \alpha' $. Then:
    \begin{enumerate}
        \item $ \tau(\alpha) = 2 $ if and only if $ \alpha' = a^\omega $, if and only if $ \alpha = a^\omega $.
        \item $ \tau(\alpha) \not = 2 $ if and only if $ \alpha' \not = a^\omega $, if and only if $ \alpha \not = a^\omega $.
    \end{enumerate}
    Assume that $ \tau(\alpha) \not = 2 $. Then, there exist unique $ c \in \Sigma \setminus \{a \} $, $ \alpha'' \in \Sigma^\omega $ and $ i \ge 0 $ such that $ \alpha' = a^i c \alpha'' $ (and so $ \alpha = a^{i + 1} c \alpha'' $). Moreover:
    \begin{enumerate}
        \item $ \tau(\alpha) = 1 $ if and only if $ c \prec a $, if and only if $ \alpha' \prec a^\omega $, if and only if $ \alpha \prec a^\omega $.
        \item $ \tau(\alpha) = 3 $ if and only if $ a \prec c $, if and only if $ a^\omega \prec \alpha' $, if and only if $ a^\omega \prec \alpha $,
    \end{enumerate}

\begin{proof}
    \begin{enumerate}
        \item Let us prove that $ \tau(\alpha) = 2 $ if and only if $ \alpha' = a^\omega $.

        $ (\Leftarrow) $ If $ \alpha' = a^\omega $, we have $ \alpha = a \alpha' = a a^\omega = a^\omega = \alpha' $, so $ \tau(\alpha) = 2 $.
        
        $ (\Rightarrow) $ Assume that $ \tau(\alpha) = 2 $ and so $ \alpha = \alpha' $. In order to prove that $ \alpha' = a^\omega $, it will suffice to show that for every $ i \ge 1 $ it holds $ \alpha' = a^i \alpha' $. We proceed by induction on $ i $. Notice that $ \alpha' = \alpha = a \alpha' $, which proves our claim for $ i = 1 $. Now, assume that $ i > 1 $. By the inductive hypothesis, we can assume $ \alpha' = a^{i - 1} \alpha' $. Hence, $ \alpha' = \alpha = a \alpha' = a (a^{i - 1} \alpha') = a^i \alpha' $.

        Let us prove that $ \alpha' = a^\omega $ if and only if $ \alpha = a^\omega $. We have $ \alpha' = a^\omega $ if and only if $ a \alpha' = a a^\omega $, if and only if $ \alpha = a^\omega $.

        \item It follows by negating the first point.
    \end{enumerate}
    Now assume that $ \tau(\alpha) \not = 2 $. Since $ \alpha' \not = a^\omega $, then there exist unique $ c \in \Sigma \setminus \{a \} $, $ \alpha'' \in \Sigma^\omega $ and $ i \ge 0 $ such that $ \alpha' = a^i c \alpha'' $. Moreover:
    \begin{enumerate}
        \item We have $ \tau (\alpha) = 1 $ if and only if $ \alpha' \prec \alpha $, if and only if $ \alpha' \prec a \alpha' $, if and only if $ a^i c \alpha'' \prec a (a^i c \alpha'') $, if and only if $ a^i c \alpha'' \prec a^i (a c \alpha'') $, if and only if $ c \alpha'' \prec a c \alpha'' $, if and only if $ c \prec a $, if and only if $ a^i c \alpha'' \prec a^\omega $, if and only if $ \alpha' \prec a^\omega $, if and only if $ a \alpha' \prec a a^\omega $, if and only if $ \alpha \prec a^\omega $.
        \item It follows by negating the previous points.
    \end{enumerate}
\end{proof}

\noindent{\textbf{Statement of Corollary \ref{cor:typesproperties}}.}
    Let $ \alpha, \beta \in \Sigma^\omega $. Let $ a, b \in \Sigma $ and $ \alpha', \beta' \in \Sigma^\omega $ such that $ \alpha = a \alpha' $ and $ \beta = b \beta'$. Then:
    \begin{enumerate}
        \item If $ a = b $ and $ \tau (\alpha) = \tau (\beta) = 2 $, then $ \alpha = \beta $.
        \item If $ a = b $ and $ \tau (\alpha) < \tau (\beta) $, then $ \alpha \prec \beta $. Equivalently, if $ a = b $ and $ \alpha \preceq \beta $, then $ \tau (\alpha) \le \tau (\beta) $.
    \end{enumerate}

\begin{proof}
    \begin{enumerate}
        \item By Lemma \ref{lem:typeequivalent} we have $ \alpha = a^\omega = b^\omega = \beta $.
        \item Let us prove that, if $ a = b $ and $ \tau (\alpha) < \tau (\beta) $, then $ \alpha \prec \beta $. We distinguish three cases.
    \begin{enumerate}
        \item $ \tau(\alpha) = 1 $ and $ \tau(\beta) = 2 $. Then by Lemma \ref{lem:typeequivalent} we have $ \alpha \prec a^\omega = b^\omega = \beta $.
        \item $ \tau(\alpha) = 2 $ and $ \tau(\beta) = 3 $. Then by Lemma \ref{lem:typeequivalent} we have $ \alpha = a^\omega = b^\omega \prec \beta $.
        \item $ \tau(\alpha) = 1 $ and $ \tau(\beta) = 3 $. Then by Lemma \ref{lem:typeequivalent} we have $ \alpha \prec a^\omega = b^\omega \prec \beta $.
    \end{enumerate}
    Lastly, if $ a = b $ and $ \alpha \preceq \beta $, then $ \tau (\alpha) \le \tau (\beta) $, because if it were $ \tau (\beta) \le \tau (\alpha) $, then we would conclude $ \beta \prec \alpha $, a contradiction.
    \end{enumerate}
\end{proof}

\section{Proofs from Section \ref{sec:computingmin}} 

\noindent{\textbf{Statement of Lemma \ref{lem:trimming}}.}
    Let $ G = (V, E) $ be a graph. Then, in $ O(|E|) $ time we can build a trimmed graph $ G^* =(V^*, E^*) $, with $ V^* = \{u^* \;|\; u \in V \} $, such that for every $ u \in V $ it holds $ \min_{u^*} = \min_{u} $. In particular, $ \tau(u^*) = \tau (u) $ for every $ u \in V $.

\begin{proof}
    Define $ \lambda (u^*) = \lambda (u) $ for every $ u^* \in V^* $, $ F = \{(u, v) \in E \;|\; \tau(v) = 1 , \lambda (v) \prec \lambda (u) \} $, and $ E^* = \{(u^*, v^*) \;|\; (u, v) \in E \setminus F \} $. Essentially, we define $ G^* $ by removing from $ G $ all edges that violate the definition of trimmed graph. 
    It is easy to realize that $ G^* $ still satisfies the assumptions on graphs that we make in this paper.
    \begin{enumerate}
        \item All edges entering the same node in $ V^* $ have the same label by construction.
        \item Every $ v^* \in G^* $ has an incoming edge in $ G^* $. Indeed, if $ \tau (v) \not = 1 $, then, if $ u \in V $ is any node such that $ (u, v) \in E $, then $ (u, v) \not \in F $ and so $ (u^*, v^*) \in E^* $. If $ \tau (v) = 1 $, then there exists $ u \in V $ such that $ (u, v) \in E $ and $ \lambda (u) \preceq \lambda (v) $, so $ (u, v) \not \in F $ and $ (u^*, v^*) \in E^* $.
        \item $ G^* $ is deterministic because it is essentially a subgraph of a deterministic graph. More precisely, assume that $ (u^*, v^*). (u^*, z^*) \in E^* $ are such that $ \lambda (v^*) = \lambda (z^*) $. We must prove that $ v^* = z^* $, or equivalently, $ v = z $. From $ (u^*, v^*), (u^*, z^*) \in E^* $ we obtain $ (u, v), (u, z) \in E $, so from $ \lambda (v) = \lambda (v^*) = \lambda (z^*) = \lambda (z) $ we obtain $ v = z $ because $ G $ is deterministic.
    \end{enumerate}
        Now, let us prove that for every $ u \in V $ it holds $ \min_{u^*} = \min_{u} $. To this end we have to prove that (i) $ \min_{u} \in I_{u^*} $ and (ii) if $ \alpha \in I_{u^*} $, then $ \min_u \preceq \alpha $.
    \begin{enumerate}
        \item Let us prove that $ \min_{u} \in I_{u^*} $. To this end, it will suffice to prove that, if $ (u_i)_{i \ge 1} $ is an occurrence of $ \min_u $ starting at $ u $, then $ (u^*_i)_{i \ge 1} $ is an occurrence of $ \min_{u} $ starting at $ u^* $. For every $ i \ge 1 $, we have $ \lambda (u^*_i) = \lambda (u_i) = \min_u [i] $. We are only left with showing that $ (u^*_{i + 1}, u^*_i) \in E^* $ for every $ i \ge 1 $. We know that $ (u_{i + 1}, u_i) \in E $, so we only have to prove that $ (u_{i + 1}, u_i) \not \in F $. Assume that $ \tau (u_i) = 1 $. We must prove that $ \lambda (u_{i + 1}) \preceq \lambda (u_i) $. Since $ (u_j)_{j \ge i} $ is an occurrence of $ \min_{u_i} $ starting at $ u_i $, this is equivalent to proving that $ \min_{u_i}[2] \preceq \min_{u_i}[1] $, which indeed follows from $ \tau (u_i) = 1 $.
        \item Let us prove that if $ \alpha \in I_{u^*} $, then $ \alpha \in I_u $ (which implies $ \min_u \preceq \alpha $). Let $ (v^*_i)_{i \ge 1} $ be an occurrence of $ \alpha $ starting at $ u^* $. It will suffice to prove that $ (v_i)_{i \ge 1} $ is an occurrence of $ \alpha $ starting at $ u $. For every $ i \ge 1 $ we have $ (v_{i + 1}, v_i) \in E $ because $ (v^*_{i + 1}, v^*_i) \in E $, and $ \lambda (v_i) = \lambda (v_i) = \alpha [i] $, so the conclusion follows.
    \end{enumerate}
    In particular, $ \tau(u^*) = \tau (u) $ for every $ u \in V $, and $ G^* =(V^*, E^*) $ is a trimmed graph.
\end{proof}

\subsection{Proofs from Section \ref{sec:minimaclassification}} 

\noindent{\textbf{Statement of Lemma \ref{lem:nextL}}.}
    Let $ G = (V, E) $ be a graph, and let $ u, v \in V $.
    \begin{enumerate}
        \item If $ (u, v) \in E $ and $ \lambda (u) \prec \lambda (v) $, then $ \tau (v) = 1 $.
        \item If $ (u, v) \in E $, $ \lambda (u) = \lambda (v) $ and $ \tau(u) = 1 $, then $ \tau(v) = 1 $.
    \end{enumerate}

\begin{proof}
    \begin{enumerate}
        \item Since $ \lambda (u) \prec \lambda (v) $, we have $ \min_u \prec \lambda (v)^\omega $. Moreover, $ \lambda (v) \min_u \in I_v $, hence $ \min_v \preceq \lambda (v) \min_u \prec \lambda (v) \lambda(v)^\omega = \lambda (v)^\omega $, so $ \tau (v) = 1 $ by Lemma \ref{lem:typeequivalent}.
        \item     Since $ \tau(u) = 1 $, then $ \min_u \prec \lambda (u)^\omega $ by Lemma \ref{lem:typeequivalent}. From $ (u, v) \in E $ we obtain $ \lambda (v) \min_u \in I_v $, hence $ \min_v \preceq \lambda (v) \min_u \prec \lambda (v) \lambda (u)^\omega = \lambda (v) \lambda (v)^\omega = \lambda (v)^\omega $, so again by Lemma \ref{lem:typeequivalent} we conclude $ \tau(v) = 1 $.
    \end{enumerate}
\end{proof}

\noindent{\textbf{Statement of Corollary \ref{cor:findingt(u)=1}}.}
    Let $ G = (V, E) $ be a graph, and let $ u \in V $. Then, $ \tau (u) = 1 $ if and only if there exist $ k \ge 2 $ and $ z_1, \dots, z_k \in V $ such that (i) $ (z_i, z_{i + 1}) \in E $ for every $ 1 \le i \le k - 1$, (ii) $ z_k = u $, (iii) $ \lambda (z_1) \prec \lambda (z_2) $ and (iv) $ \lambda (z_2) = \lambda (z_3) = \dots = \lambda (z_k) $.

\begin{proof}
    $ (\Leftarrow) $ Let $ k \ge 2 $ and $ z_1, \dots, z_k \in V $ be such that (i) $ (z_i, z_{i + 1}) \in E $ for every $ 1 \le i \le k - 1$, (ii) $ z_k = u $, (iii) $ \lambda (z_1) \prec \lambda (z_2) $ and (iv) $ \lambda (z_2) = \lambda (z_3) = \dots = \lambda (z_k) $. We must prove that $ \tau(u) = 1 $. Let us prove by induction that for every $ 2 \le i \le k $ it holds $ \tau(z_i) = 1 $ (and in particular $ \tau(u) = \tau (z_k) = 1 $). If $ i = 2 $, then from $ (z_1, z_2) \in E $ and $ \lambda (z_1) \prec \lambda (z_2) $ we obtain $ \tau(z_2) = 1 $ by Lemma \ref{lem:nextL}. Now, assume that $ 3 \le i \le k $. By the inductive hypothesis, we have $ \tau(z_{i - 1}) = 1 $. Since $ (z_{i - 1}, z_i) \in E $ and $ \lambda (z_{i -1}) = \lambda (z_i) $, then by Lemma \ref{lem:nextL} we conclude $ \tau(z_i) = 1 $.

    $ (\Rightarrow) $ Assume that $ \tau(u) = 1 $. Then, by Lemma \ref{lem:typeequivalent} there exist $ c \in \Sigma \setminus \{\lambda (u) \} $, $ \gamma' \in \Sigma^\omega $ and $ j \ge 1 $ such that $ \min_u = \lambda(u)^j c \gamma'$ and $ c \prec \lambda (u) $. Let $ (u_i)_{i \ge 1} $ be an occurrence of $ \min_u $ starting at $ u $. Then, (i) $ u = u_1 $, (ii) $ (u_{i + 1}, u_i) \in E $ for every $ i \ge 1 $, (iii) $ \lambda (u_i) = \lambda (u) $ for $ 1 \le i \le j $ and (iv) $ \lambda (u_{j + 1}) = c $. As a consequence, if $ k = j + 1 $ and $ z_i = u_{k + 1 - i} $ for every $ 1 \le i \le k $, then (i) $ (z_i, z_{i + 1}) = (u_{k - i + 1}, u_{k - i}) \in E $ for every $ 1 \le i \le k - 1$, (ii) $ z_k = u_1 = u $, (iii) $ \lambda (z_1) = \lambda (u_k) = \lambda (u_{j + 1}) = c \prec \lambda (u) = \lambda (u_j) = \lambda(u_{k - 1}) = \lambda (z_2) $ and (iv) $ \lambda (z_2) = \lambda (z_3) = \dots = \lambda (z_k) $ because $ \lambda (u_j) = \lambda (u_{j - 1}) = \dots = \lambda (u_1) $.
\end{proof}

\noindent{\textbf{Statement of Corollary \ref{cor:type1}}.}
    Let $ G = (V, E) $ be a graph. We can determine all $ u \in V $ such that $ \tau (u) = 1 $ in time $ O(|E|) $.

\begin{proof}
    In our algorithm, we will mark exactly all nodes $ u $ such that $ \tau (u) = 1 $ by means of a graph traversal. We will use an (initially empty) queue. Initially, no node is marked. First, process all edges $ (u, v) \in E $ and, if $ \lambda (u) \prec \lambda (v) $ and $ v $ has not been marked before, then mark $ v $ and add $ v $ to the queue. This step takes $ O(m) $ time. Next, recursively pick an element $ u $ from the queue, and consider the unique $ v \in Q $ such that $ (u, v) \in E $ and $ \lambda (u) = \lambda (v) $ (if it exists). If $ v $ has not been marked before, then mark $ v $ and add $ v $ to the queue. This step also takes $ O(|E|) $ time because each edge is considered at most once. At the end of the algorithm, by Corollary \ref{cor:findingt(u)=1} a node $ u $ has been marked if and only if $ \tau (u) = 1 $.
\end{proof}

\noindent{\textbf{Statement of Lemma \ref{lem:characterizationtype2}}.}
     Let $ G = (V, E) $ be a graph, and let $ u \in V $ such that $ \tau (u) \not = 1 $. Then, we have $ \tau (u) = 2 $ if and only if there exist $ k \ge 2 $ and $ z_1, \dots, z_k \in V $ such that (i) $ (z_{i + 1}, z_i) \in E $ for every $ 1 \le i \le k - 1 $, (ii) $ z_1 = u $, (iii) $ z_k = z_j $ for some $ 1 \le j \le k - 1 $ and (iv) $ \lambda (z_1) = \lambda (z_2) = \dots = \lambda (z_k) $.

     In particular, such $ z_1, \dots, z_k \in V $ must satisfy $ \tau (z_i) = 2 $ for every $ 1 \le i \le k $.

\begin{proof}
    $ (\Leftarrow) $ Let $ k \ge 2 $ and $ z_1, \dots, z_k \in V $ such that (i) $ (z_{i + 1}, z_i) \in E $ for every $ 1 \le i \le k - 1 $, (ii) $ z_1 = u $, (iii) $ z_k = z_i $ for some $ 1 \le j \le k - 1 $ and (iv) $ \lambda (z_1) = \lambda (z_2) = \dots = \lambda (z_k) $. We must prove that $ \tau (u) = 2 $. Notice that $ \min_{u} = \min_{z_1} \preceq \lambda (z_1) \lambda (z_2) \dots \lambda (z_{j - 1}) \min_{z_j} = \lambda (u)^{j - 1} \min_{z_j} $. Similarly, we have $ \min_{z_j} \preceq \lambda (z_j) \lambda (z_{j + 1}) \dots \lambda (z_{k - 1}) \min_{z_k} = \lambda (u)^{k - j} \min_{z_j} $. Since $ \min_{z_j} \preceq \lambda (u)^{k - i} \min_{z_j} $, then by induction we obtain $ \min_{z_i} \preceq (\lambda (u)^{k - j})^h \min_{z_i} $ for every $ h \ge 1 $, and so $ \min_{z_j} \preceq \lambda (u)^\omega $. As a consequence, $ \min_{u} \preceq \lambda (u)^{j - 1} \min_{z_j} \preceq \lambda (u)^{j - 1} \lambda (u)^\omega = \lambda (u)^\omega $, and so $ \tau(u) \preceq 2 $ by Lemma \ref{lem:typeequivalent}. Since $ \tau(u) \not = 1 $, we conclude $ \tau (u) = 2 $.

    $ (\Rightarrow) $ Assume that $ \tau(u) = 2 $. In particular, $ \lambda (u)^\omega \in I_u $, so there exists an occurrence $ (z_i)_{i \ge 1} $ of $ \lambda (u)^\omega $ starting at $ u $. This means that (i) $ (z_{i + 1}, z_i) \in E $ for every $ i \ge 1 $, (ii) $ z_1 = u $ and $ \lambda (z_i) = \lambda (u) $ for every $ i \ge 1 $. Since $ V $ is finite, there exist $ 1 \le j < k $ such that $ z_j = z_k $. Then, $ k \ge 2 $ and $ z_1, \dots, z_k \in V $ have the desired properties.

    Now, let us prove that, if there exist $ z_1, \dots, z_k \in V $ with the desired properties, then $ \tau (z_i) = 2 $ for every $ 1 \le i \le k $. Notice that it must be $ \tau (z_i) \not = 1 $ for every $ 1 \le i \le k $, because otherwise by Corollary \ref{cor:findingt(u)=1} we would conclude $ \tau (u) = 1 $. As a consequence, it must be $ \tau (z_i) = 2 $ for every $ 1 \le i \le k $ by the characterization that we have just proved.
\end{proof}

\noindent{\textbf{Statement of Corollary \ref{cor:type2}}.}
     Let $ G = (V, E) $ be a graph. We can determine all $ u \in V $ such that $ \tau (u) = 2 $ in time $ O(|E|) $.

\begin{proof}
    By Corollary \ref{cor:type1}, we can assume that we have already computed all $ u \in V $  such that $ \tau (u) = 1 $. In order to compute all $ u \in V $ such that $ \tau (u) = 2 $, we explore $ G $ by using a breadth first search algorithm, with the following modifications: (i) we only start from nodes $ u \in V $ such that $ \tau(u) \not = 1 $, (ii) we follow edges in a backward fashion, not in a forward fashion and (iii) if we start from $ u \in V $, we explore the graph by only following edges labeled $ \lambda (u) $ (that is, we first consider only all $ u' \in V $ such that $ (u', u) \in E $ and $ \lambda (u') = \lambda (u) $, then we repeat this step in BFS fashion). Note that during the search we can never encounter a node $ v $ such that $ \tau (v) = 1 $, otherwise by Corollary \ref{cor:findingt(u)=1} we would obtain $ \tau(u) = 1 $. During the search, we will infer the value $ \tau (v) $ for every node $ v $ that we encounter. Assume that we start the exploration from node $ u \in V $. If during the exploration at $ u $ at some point we encounter a node $ v $ for which we have already concluded that $ \tau (v) = 3 $, then we do not consider the edges entering $ v $ and we backtrack (because all the $ z_i $'s in the characterization of Lemma \ref{lem:characterizationtype2} must satisfy $ \tau (z_i) = 2 $). If during the exploration at $ u $ at some point we encounter a node $ v $ for which we have already concluded that $ \tau (v) = 2 $, then we backtrack straight to $ u $ and, by Lemma \ref{lem:characterizationtype2}, we can also conclude that all $ z $ that we encounter during the backtracking (including $ u $) are such that $ \tau (z) = 2 $. If during the exploration of $ u $ at some point we encounter the same node twice, then we backtrack straight to $ u $ and, by Lemma \ref{lem:characterizationtype2}, we can also conclude that all $ z $ that we encounter during the backtracking (including $ u $) are such that $ \tau (z) = 2 $. If during the exploration of $ u $ we never encounter a node $ v $ for which we have already determined $ \tau(v) $ and  we never encounter the same node twice (and so at some point we get stuck), by Lemma \ref{lem:characterizationtype2} we can conclude that all $ v $ that we have encountered (including $ u $) are such that $ \tau (v) = 3 $. The algorithm runs in $ O(|E|) $ time because we never follow the same edge twice.
\end{proof}

\subsection{Proofs from Section \ref{sec:recursivestep}} 

\noindent{\textbf{Statement of Lemma \ref{lem:comparingminima}}.}
    Let $ G = (V, E) $ be a graph. Let $ u, v \in V $ be such that $ \min_u = \min_v $. Let $ (u_i)_{i \ge 1} $ be an occurrence of $ \min_u $ and let $ (v_i)_{i \ge 1} $ be an occurrence of $ \min_v $. Then:
    \begin{enumerate}
        \item $ \lambda (u_i) = \lambda (v_i) $ for every $ i \ge 1 $.
        \item $ \min_{u_i} = \min_{v_i} $ for every $ i \ge 1 $.
        \item $ \tau(u_i) = \tau(v_i) $ for every $ i \ge 1 $.
    \end{enumerate}
    In particular, the previous results hold if $ u = v $ and $ (u_i)_{i \ge 1} $ and $ (v_i)_{i \ge 1} $ are two distinct occurrences of $ \min_u $.

\begin{proof}
    \begin{enumerate}
        \item For every $ i \ge 1 $ we have $ \lambda (u_i) = \min_u[i] = \min_v[i] = \lambda (v_i) $.
        \item Fix $ i \ge 1 $. We have $ \min_u[1, i - 1] \min_{u_i} = \min_u = \min_v = \min_v[1, i - 1] \min_{v_i} $ and so $ \min_{u_i} = \min_{v_i} $
        \item Fix $ i \ge 1 $. By the previous points we have $ \min_{u_i} = \min_{v_i} $ and $ \lambda (u_i) = \lambda (v_i) $, so by Corollary \ref{cor:typesproperties} we conclude that it must necessarily be $ \tau(u_i) = \tau(v_i) $.
    \end{enumerate}
\end{proof}

\noindent{\textbf{Statement of Lemma \ref{lem:minpairwisedistinct}}.}
Let $ G = (V, E) $ be a graph. Let $ u \in V $ and let $ (u_i)_{i \ge 1} $ an occurrence of $ \min_u $ starting at $ u $. Let $ k \ge 1 $ be such that $ \tau (u_1) = \tau (u_2) = \dots = \tau (u_{k - 1}) = \tau (u_k) \not = 2 $. Then, $ u_1, \dots, u_k $ are pairwise distinct. In particular, $ k \le |V| $.

\begin{proof}
    We assume that $ \tau (u_1) = \tau (u_2) = \dots = \tau (u_{k - 1}) = \tau (u_k) = 1 $, because the case $ \tau (u_1) = \tau (u_2) = \dots = \tau (u_{k - 1}) = \tau (u_k) = 3 $ is symmetrical. Notice that for every $ 1 \le l \le k $ we have that $ (u_i)_{i \ge l} $ is an occurrence of $ \min_{u_l} $ starting at $ u_l $ and $ \tau (u_l) = 1 $, so $ \lambda (u_{l + 1}) \preceq \lambda (u_l) $. In particular, for every $ 1 \le r \le s \le k + 1 $ we have $ \lambda (u_s) \preceq \lambda(u_r) $.
    
    Suppose for sake of contradiction that there exist $ 1 \le i < j \le k $ such that $ u_i = u_j $. Then, $ \min_{u_i} = \min_{u_i}[1, j - i] \min_{u_j} = \min_{u_i}[1, j - i] \min_{u_i} $. By induction, we obtain $ \min_{u_i} = (\min_{u_i}[1, j - i])^t \min_{u_i} $ for every $ t \ge 1 $, and so $ \min_{u_i} = (\min_{u_i}[1, j - i])^\omega $. For every $ 1 \le h \le j $, we have $ \lambda (u_j) \preceq \lambda (u_h) \preceq \lambda (u_i) $. Since $ u_i = u_j $, then $ \lambda (u_i) = \lambda (u_j) $, so $ \lambda (u_h) = \lambda (u_i) $ for every $ i \le h \le j $, which implies $ \min_{u_i}[1, j - i] = \lambda (u_i) \lambda (u_{i + 1}) \dots \lambda (u_{j - 1}) = \lambda (u_i)^{j - i} $ and so $ \min_{u_i} = (\min_{u_i}[1, j - i])^\omega = \lambda (u_i)^\omega $. By Lemma \ref{lem:typeequivalent} we conclude $ \tau (u_i) = 2 $, a contradiction.
\end{proof}

\noindent{\textbf{Statement of Lemma \ref{lem:charatersdecrease}}.}
    Let $ G = (V, E) $ be a graph. Let $ u \in V $ such that $ \tau (u) = 3 $. Then, $ \min_u [i + 1] \preceq \min_u[i] $ for every $ 2 \le i \le \ell_u - 1 $. In particular, if $ 2 \le i \le j \le \ell_u $, then $ \min_u [j] \preceq \min_u [i] $.

\begin{proof}
    Fix $ 2 \le i \le \ell_u $. By the definition of $ \ell_u)$ we have $ \tau (u_i) = 1 $, so $ \min_u [i + 1] = \min_{u_i}[2] \preceq \min_{u_i}[1] = \min_u[i] $.
\end{proof}

\noindent{\textbf{Statement of Lemma \ref{lem:propertiesexploringminimum}}.}
    Let $ G = (V, E ) $ be a graph. Let $ u \in V $ such that $ \tau (u) = 3 $.
    \begin{enumerate}
        \item $ G_i (u) $ is well-defined and nonempty for every $ 1 \le i \le \ell_u $.
        \item Let $ (u_i)_{i \ge 1} $ be an occurrence of $ \min_u $ starting at $ u $. Then, $ u_i \in G_i (u) $ for every $ 1 \le i \le \ell_u $. In particular, $ \tau(u_i) = \tau(G_i(u))$ and $ \min_u[i] = \lambda(u_i) = \lambda(G_i(u))$ for every $ 1 \le i \le \ell_u $.
        \item For every $ 1 \le i \le \ell_u $ and for every $ v \in G_i (u) $ there exists an occurrence of $ \min_u[1, i - 1] $ starting at $ u $ and ending at $ v $.
    \end{enumerate}

\begin{proof}
    We proceed by induction on $ i \ge 1 $. If $ i = 1 $, then $ G_i (u) = \{u \} $ is well-defined and nonempty, and $ u_1 = u \in G_i (u) $. Moreover, if $ v \in G_i (u) $, then $ v = u $, and there exists an occurrence of $ \epsilon = \min [1, 0] $ starting and ending at $ u $. Now, assume that $ 2 \le i \le \ell_u $.

    First, notice that $ \mathtt{F}(\{v' \in Q \;|\; (\exists v \in G_{i - 1}(u))((v', v) \in E)\}) $ is well-defined and nonempty. Indeed, by the inductive hypothesis $ G_{i - 1} (u) $ is well-defined and nonempty, so $ \{v' \in Q \;|\; (\exists v \in G_{i - 1}(u))((v', v) \in E)\} $ is nonempty because every node has an incoming edge.

    Let us prove that $ u_i \in \mathtt{F}(\{v' \in Q \;|\; (\exists v \in G_{i - 1}(u))((v', v) \in E)\}) $. Let $ R = \{v' \in Q \;|\; (\exists v \in G_{i - 1}(u))((v', v) \in E)\} $ and $ R' = \argmin_{v \in R} \lambda (v) $. We must prove that $ u_i \in \argmin_{u \in R'} \tau(u) $.
    \begin{enumerate}
        \item Let us prove that $ u_i \in R $. By the inductive hypothesis, we know that $ u_{i - 1} \in G_{i - 1}(u) $. We know that $ (u_i, u_{i - 1}) \in E $, so $ u_i \in R $.
        \item Let us prove that $ u_i \in R' $. Let $ v' \in R $. We must prove that $ \lambda (u_i) \preceq \lambda (v') $. Since $ v' \in R $, there exists $ v \in G_{i - 1}(u) $ such that $ (v', v) \in E $. From $ u_{i - 1}, v \in G_{i - 1}(u) $ we obtain $ \lambda (u_{i - 1}) = \lambda (v) $. By the inductive hypothesis, there exists an occurrence of $ \min_u[1, i - 2] $ starting at $ u $ and ending at $ v $, so there exists an occurrence of $ \min_u[1, i - 2] \lambda (v) = \min_u[1, i - 2] \lambda (u_{i - 1}) = \min_u[1, i - 1] $ starting at $ u $ and ending at $ v' $. Since $ \min_u = \min_u[1, i - 1] \min_{u_i} $, it must be $ \min_{u_i} \preceq \min_{v'} $, and in particular $ \lambda (u_i) \preceq \lambda (v') $.
        \item Let us prove that $ u_i \in \argmin_{u \in R'} \tau(u) $. Let $ v' \in R' $. We must prove that $ \tau (u_i) \le \tau (v) $. In particular, $ v' \in R $, so as before we obtain $ \min_{u_i} \preceq \min_{v'} $. Moreover, from $ u_i, v' \in R' $ we obtain $ \lambda (u_i) = \lambda (v') $. From Corollary \ref{cor:typesproperties} we conclude $ \tau (u_i) \le \tau (v) $.
    \end{enumerate}

    Let us prove that $ u_i \not \in \bigcup_{j = 2}^{i - 1} G_j (u) $. If $ i = \ell_u $, we have $ \tau (u_i) = \tau(u_{\ell_u}) \ge 2 $ and the conclusion follows because $ \tau (G_j (u)) = \tau(u_j) = 1 $ for every $ 2 \le j \le i - 1 $. As a consequence, in the following we can assume $ 2 \le i \le \ell_u - 1 $. In particular, $ \tau (u_i) = 1 $, so by Lemma \ref{lem:typeequivalent} there exists $ k \ge 1 $, $ c \in \Sigma \setminus \{\lambda (u_i) \} $ and $ \gamma' \in \Sigma^\omega $ such that $ \min_u = \lambda (u)^k c \gamma' $ and $ c \prec \lambda (u_i) $. Suppose for sake of contradiction that there exists $ 2 \le j \le i - 1 \le \ell_u - 2 $ such that $ u_i \in G_j (u) $. We know that $ u_j \in G_j (u) $, and in particular, $ \lambda (u_i) = \lambda (G_j (u)) = \lambda (u_j) $. From Lemma \ref{lem:charatersdecrease} we obtain $ \min_u[i] \preceq \min_u[h] \preceq \min_u[j] $ for every $ j \le h \le i $, or equivalently $ \lambda (u_i) \preceq \lambda (u_h) \preceq \lambda(u_j) $ for every $ j \le h \le i $. Since $ \lambda (u_j) = \lambda (u_i) $, we conclude $ \lambda (u_h) = \lambda (u_i) $ for every $ j \le h \le i $. As a consequence, $ \min_u = \min_u [1, i - 1] \min_{u_i} = \min_u [1, j - 1] \min_u[j, i - 1] \min_{u_i} = \min_u [1, j - 1] \lambda (u_i)^{i - j} \lambda (u_i)^k c \gamma' = \min_u [1, j - 1] \lambda (u_i)^{k + i - j} c \gamma' $. On the other hand, since $ u_i \in G_j (u) $ and $ j < i $, by the inductive hypothesis there exists an occurrence of $ \min_u[1, j - 1] $ starting at $ u $ and ending at $ u_i $, so $ \min_u[1, j - 1] \min_{u_i} \in I_u $. As a consequence, by the minimality of $ \min_u $ we obtain $ \min_u \preceq \min_u[1, j - 1] \min_{u_i} $, or equivalently, $ \min_u[1, j - 1] \lambda (u_i)^{k + i - j} c \gamma' \preceq \min_u[1, j - 1] \lambda (u_i)^{k} c \gamma' $. Since $ j < i $, we obtain $ \lambda (u_i) \preceq c $, a contradiction.

    Let us prove that $ G_i (u) $ is well-defined and nonempty, and $ u_i \in G_i (u) $. This follows from $ u_i \in \mathtt{F}(\{v' \in Q \;|\; (\exists v \in G_{i - 1}(u))((v', v) \in E)\}) $ and $ u_i \not \in \bigcup_{j = 2}^{i - 1} G_j (u) $.

    Lastly, let us prove that if $ v' \in G_i (u) $, then there exists an occurrence of $ \min[1, i - 1] $ starting at $ u $ and ending at $ v' $. Since $ v' \in G_i (u) $, then there exists $ v \in G_{i - 1}(u) $ such that $ (v', v) \in E $. In particular, $ \lambda (v) = \lambda (G_{i - 1}(u)) = \lambda (u_{i - 1}) $. By the inductive hypothesis, there exists an occurrence of $ \min_u[1, i - 2] $ starting at $ u $ and ending at $ v $, so there exists an occurrence of $ \min_u[1, i - 2] \lambda (v) = \min_u[1, i - 2] \lambda (u_{i - 1}) = \min_u[1, i - 1] $ starting at $ u $ and ending at $ v' $.
\end{proof}

\noindent{\textbf{Statement of Lemma \ref{lem_motivatingorder}}.}
    Let $ G = (V, E) $ be a graph. Let $ u, v \in V $ such that $ \tau(u) = \tau(v) = 3 $. Assume that one of the following statements is true:
    \begin{enumerate}
        \item $ \gamma_u $ is not a prefix of $ \gamma_v $ and $ \gamma_u \prec \gamma_v $.
        \item $ \gamma_u = \gamma_v $, $ \mathtt{t}_u = 2 $ and $ \mathtt{t}_v = 3 $.
        \item $ \gamma_v $ is a strict prefix of $ \gamma_u $.
    \end{enumerate}
    Then, $ \min_u \prec \min_v $.
    
    Equivalently, if $ \min_u \preceq \min_v $, then one the following is true: (i) $ \gamma_u $ is not a prefix of $ \gamma_v $ and $ \gamma_u \prec \gamma_v $; (ii) $ \gamma_u = \gamma_v $ and $ \mathtt{t}_u \le \mathtt{t}_v $; (iii) $ \gamma_v $ is a strict prefix of $ \gamma_u $.

\begin{proof}
    Let us prove that, if one of the three statements (1) - (3) is true, then $ \min_u \prec \min_v $.
    \begin{enumerate}
        \item If $ \min_u[1, \ell_u] \prec \min_v[1, \ell_v] $ and $ \min_u[1, \ell_u] $ is not a prefix of $ \min_v[1, \ell_v] $, then $ \min_u \prec \min_v $.
        \item Assume that $ \min_u[1, \ell_u] = \min_v[1, \ell_v] $ (so in particular $ \ell_u = \ell_v $), $ \mathtt{t}_u = 2 $ and $ \mathtt{t}_v = 3 $. Let $ (u_i)_{i \ge 1} $ be an occurrence of $ u $, and let $ (v_i)_{i \ge 1} $ be an occurrence of $ v $. It holds $ \min_u = \min_u[1, \ell_u - 1] \min_{u_{\ell_u}} $ and $ \min_v = \min_v[1, \ell_v - 1] \min_{v_{\ell_v}} $. Since $ \min_u[1, \ell_u - 1] = \min_v[1, \ell_v - 1] $, in order to prove that $ \min_u \prec \min_v $ we only have to show that $ \min_{u_{\ell_u}} \prec \min_{v_{\ell_v}} $. By Lemma \ref{lem:propertiesexploringminimum} we have $ u_{\ell_u} \in G_{\ell_u}(u) $ and $ v_{\ell_v} \in G_{\ell_v}(v) $, so $ \lambda (u_{\ell_u}) = \min_u[\ell_u] = \min_v[\ell_v] = \lambda (v_{\ell_v}) $. Since $ \tau(u_{\ell_u}) = \tau (G_{\ell_u} (u)) = \mathtt{t}_u = 2 $ and $ \tau(v_{\ell_v}) = \tau (G_{\ell_v} (v)) = \mathtt{t}_v = 3 $, we conclude $ \min_{u_{\ell_u}} \prec \min_{v_{\ell_v}} $ by Corollary \ref{cor:typesproperties}.
        \item Assume that $ \gamma_v $ is a strict prefix of $ \gamma_u $ (hence $ \ell_v < \ell_u $). In particular, $ \min_u[1, \ell_v] = \min_v[1, \ell_v] $, so similarly to the previous case we only have to prove that $ \min_{u_{\ell_v}} \prec \min_{v_{\ell_v}} $. Again, we obtain $ u_{\ell_v} \in G_{\ell_v}(u) $, $ v_{\ell_v} \in G_{\ell_u}(v) $, $ \lambda (u_{\ell_v}) = \lambda (v_{\ell_v}) $ and $ \tau(v_{\ell_v}) = 3 $. Since $ \ell_v < \ell_u $, the minimality of $ \ell_u $ implies $ \tau (u_{\ell_v}) = \tau(G_{\ell_v}(u)) = 1 $. By Corollary \ref{cor:typesproperties} we conclude $ \min_{u_{\ell_v}} \prec \min_{v_{\ell_v}} $.
    \end{enumerate}
    Now, let us prove that if $ \min_u \preceq \min_v $, then one the following is true: (i) $ \gamma_u $ is not a prefix of $ \gamma_v $ and $ \gamma_u \prec \gamma_v $; (ii) $ \gamma_u = \gamma_v $ and $ \mathtt{t}_u \le \mathtt{t}_v $; (iii) $ \gamma_v $ is a strict prefix of $ \gamma_u $. If this were not true, then one of the following would be true: (1) $ \gamma_v $ is not a prefix of $ \gamma_u $ and $ \gamma_v \prec \gamma_u $; (2) $ \gamma_v = \gamma_u $, $ \mathtt{t}_v = 2 $ and $ \mathtt{t}_u = 3 $; (3) $ \gamma_u $ is a strict prefix of $ \gamma_v $. We would then conclude $ \min_v \prec \min_u $, a contradiction.
\end{proof}

\noindent{\textbf{Statement of Lemma \ref{lem:type2equalityminima}}.}
    Let $ G = (V, E) $ be a graph. Let $ u, v \in V $ be such that $ \tau(u) = \tau(v) = 3 $, $ \gamma_u = \gamma_v $ and $ \mathtt{t}_u = \mathtt{t}_v = 2 $. Then, $ \min_u = \min_v $.

\begin{proof}
    First, note that $ \gamma_u = \gamma_v $ implies $ \ell_u = \ell_v $, and it is equivalent to $ \min_u[1, \ell_u] = \min_v[1, \ell_v] $. Now, let $ (u_i)_{i \ge 1} $ be an occurrence of $ \min_u $ starting at $ u $, and let $ (v_i)_{i \ge 1} $ be an occurrence of $ \min_v $ starting at $ v $. We have $ \min_u = \min_u [1, \ell_u - 1] \min_{u_{\ell_u}} $ and $ \min_v = \min_v [1, \ell_v - 1] \min_{v_{\ell_v}} $. Since $ \min_u[1, \ell_u - 1] = \min_v[1, \ell_v - 1] $, we only have to show that $ \min_{u_{\ell_u}} = \min_{v_{\ell_v}} $. Notice that  $ \lambda (u_{\ell_u}) = \min_u[\ell_u] = \min_v[\ell_v] = \lambda (v_{\ell_v}) $, and by Lemma \ref {lem:propertiesexploringminimum} we have $ \tau(u_{\ell_u}) = \tau (G_{\ell_u}(u)) = \mathtt{t}_u = 2 $ and $ \tau(v_{\ell_v}) = \tau (G_{\ell_v}(v)) = \mathtt{t}_v = 2 $. By Corollary \ref{cor:typesproperties} we conclude $ \min_{u_{\ell_u}} = \min_{v_{\ell_v}} $.
\end{proof}

\noindent{\textbf{Statement of Lemma \ref{lem:descrptionminimum}}.}
    Let $ G = (V, E) $ be a graph. Let $ u \in V $, and let $ (u_i)_{i \ge 1} $ be an occurrence of $ \min_u $ starting at $ u $. Then, exactly one of the following holds true:
    \begin{enumerate}
        \item There exists $ i_0 \ge 1 $ such that $ \tau(u_i) \not = 2 $ for every $ 1 \le i < i_0 $ and $ \tau(u_i) = 2 $ for every $ i \ge i_0 $.
        \item $ \tau(u_i) \not = 2 $ for every $ i \ge 1 $, and both $ \tau(u_i) = 1 $ and $ \tau(u_i) = 3 $ are true for infinitely many $ i $'s.
    \end{enumerate}

\begin{proof}
    If $ \tau(u_i) \not = 2 $ for every $ i \ge 1 $, then both $ \tau(u_i) = 1 $ and $ \tau(u_i) = 3 $ are true for infinitely many $ i $'s, because otherwise we could choose $ j \ge 1 $ such that either $ \tau (u_i) = 1 $ for every $ i \ge j $ or $ \tau (u_i) = 3 $ for every $ i \ge j $, and in both cases Lemma \ref{lem:minpairwisedistinct} leads to a contradiction, because $ (u_j)_{j \ge i} $ is an occurrence of $ \min_{u_j} $ starting at $ u_j $ and $ V $ is a finite set.

    Now, assume that there exists $ i_0 \ge 1 $ such that $ \tau(u_{i_0})  = 2 $; without loss of generality, we can assume that $ i_0 $ is the smallest integer with this property. We only have to prove that if $ i \ge i_0 $, then $ \tau (u_i) = 2 $. Since $ \tau (u_{i_0}) = 2 $, then $ \min_{u_{i_0}} = (\lambda (u_{i_0}))^\omega $ by Lemma \ref{lem:typeequivalent}. Since $ (u_j)_{j \ge i_0} $ in an occurrence of $ \min_{u_{i_0}} $ starting at $ i_0 $, then $ \lambda (u_j) = \lambda (u_{i_0}) $ for every $ j \ge i_0 $, so $ \min_{u_i} = (\lambda (u_i))^\omega $ for every $ i \ge i_0 $, $ (u_j)_{j \ge i} $ being an occurrence of $ \min_{u_i} $ starting at $ u_i $. By Lemma \ref{lem:typeequivalent}, we conclude $ \tau (u_i) = 2 $ for every $ i \ge i_0 $.
\end{proof}

\noindent{\textbf{Statement of Lemma \ref{ref:minimumquotientgraph}}.}
Let $ G = (V, E) $ be a graph. Let $ u \in V $ such that $ \tau(u) = 3 $. Let $ (u_i)_{i \ge 1} $ be an occurrence of $ \min_u $ starting at $ u $. Let $ (u'_i)_{i \ge 1} $ be the infinite sequence of nodes in $ V $ obtained as follows. Consider $ L = \{k \ge 1 \;|\; \tau(u_k) = 3 \} $, and for every $ i \ge 1 $, let $ j_i \ge 1 $ be the $ i^{\text{th}} $ smallest element of $ L $, if it exists. For every $ i \ge 1 $ such that $ j_i $ is defined, let $ u'_i = u_{j_i} $, and if $ i \ge 1 $ is such that $ j_i $ is not defined (so $ L $ is a finite set), let $ u'_i = u'_{|L|} $.
Then, $ (\bar{u}'_i)_{i \ge 1} $ is an occurrence of $ \min_{\bar{u}} $ starting at $ \bar{u} $ in $ \bar{G} = (\bar{V}, \bar{E}) $.

\begin{proof}
    First, notice that $ \bar{u}'_i \in \bar{V} $ for every $ i \ge 1 $ because $ \tau(u'_i) = 3 $, and $ u'_1 = u_1 = u $ (so $ \bar{u}'_1 = \bar{u}_1 = \bar{u} $). Now, let us prove that $ (\bar{u}'_{i + 1}, \bar{u}'_i) \in \bar{E} $ for every $ i \ge 1 $. Fix $ i \ge 1 $. We distinguish three cases.
    \begin{enumerate}
        \item $ j_i $ and $ j_{i + 1} $ are both defined. This means that $ \tau(u'_i) = \tau(u'_{i + 1}) = 3 $, and $ \tau(u_k) \not = 3 $ for every $ j_i < k < j_{i + 1} $. By Lemma \ref{lem:descrptionminimum}, it must be $ \tau(u_k) = 1 $ for every $ j_i < k < j_{i + 1} $. Since $ (u_k)_{k \ge j_i} $ is an occurrence of $ \min_{u'_i} $ starting at $ u'_i $, by Lemma \ref{lem:propertiesexploringminimum} we obtain $ \mathtt{t}_{u'_i} = 3 $ and $ u'_{i + 1} \in G_{\ell_{u'_i}}(u'_i) $, so $ (\bar{u}'_{i + 1}, \bar{u}'_i) \in \bar{E} $.
        \item $ j_i $ is defined and $ j_{i + 1} $ is not defined. This means that $ i = |L| $, $ \tau(u_{j_{|L|}}) = 3 $, and by Lemma \ref{lem:descrptionminimum} there exists $ h > j_{|L|} $ such that $ \tau(u_h) = 2 $ and $ \tau(u_k) = 1 $ for every $ j_{|L|} < k < h $. Since $ (u_k)_{k \ge j_{|L|}} $ is an occurrence of $ \min_{u'_{|L|}} $ starting at $ u'_{|L|} $, by Lemma \ref{lem:propertiesexploringminimum} we obtain  $ \mathtt{t}_{u'_{|L|}} = 2 $, so $ (\bar{u}'_{i + 1}, \bar{u}'_i) = (\bar{u}'_{|L|}, \bar{u}'_{|L|}) \in \bar{E} $.
        \item $ j_i $ and $ j_{i + 1} $ are both non-defined. Then, by the previous case we conclude $ (\bar{u}'_{i + 1}, \bar{u}'_i) = (\bar{u}'_{|L|}, \bar{u}'_{|L|}) \in \bar{E} $.
    \end{enumerate}
    We are left with proving that $ \min_{\bar{u}}[i] = \lambda(\bar{u}'_i) $ for every $ i \ge 1 $. Let $ \alpha \in (\Sigma')^\omega $ be such that $ \alpha[i] = \lambda(\bar{u}'_i) $ for every $ i \ge 1 $. We have to prove that $ \alpha = \min_{\bar{u}} $. Since $ (\bar{u}'_{i + 1}, \bar{u}'_i) \in \bar{E} $ for every $ i \ge 1 $, we have $ \alpha \in I_{\bar{u}} $, and $ (\bar{u}'_i)_{i \ge 1} $ is an occurrence of $ \alpha $ starting at $ \bar{u} $. The conclusion follows if we prove that for every $ \beta \in I_{\bar{u}} $ we have $ \alpha \preceq' \beta $. Fix $ \beta \in I_{\bar{u}} $; it will suffice to show that, for every $ k \ge 1 $, if $ \alpha[1, k - 1] = \beta[1, k - 1] $, then $ \alpha [k] \preceq' \beta [k] $. Fix $ k \ge 1 $, and let $ (\bar{v}_i)_{i \ge 1} $ be an occurrence of $ \beta $ starting at $ \bar{u} $. Notice that for every $ 1 \le h \le k - 1 $ we have $ \alpha[h] = \beta[h] $, or equivalently, $ \lambda(\bar{u}'_h) = \lambda(\bar{v}_h) $, or equivalently, $ \gamma_{u'_h} = \gamma_{v_h} $ (so $ \ell_{u'_h} = \ell_{v_h} $) and $ \mathtt{t}_{u'_h} = \mathtt{t}_{v_h} $. Note that $ (\bar{v}_{i + 1}, \bar{v}_i) \in \bar{E} $ for every $ i \ge  1 $. We distinguish two cases:
    \begin{enumerate}
        \item There exists $ 1 \le h \le k - 1 $ such that $ \mathtt{t}_{u'_h} = \mathtt{t}_{v_h} = 2 $. In this case, the definition of $ u'_h $ implies $ u'_h = u'_{h + 1} = u'_{h + 2} = \dots $, and in particular $ u'_h = u'_k$. Moreover, since $ \mathtt{t}_{v_h} = 2 $ and $ (\bar{v}_{i + 1}, \bar{v}_i) \in \bar{E} $, the definition of $ \bar{G} = (\bar{V}, \bar{E}) $ implies $ v_h = v_{h + 1} = v_{h + 2} = \dots $, and in particular $ v_h = v_k $. We conclude $ \alpha[k] = \lambda (u'_k) = \lambda (u'_h) = \lambda (v_h) = \lambda (v_k) = \beta [k] $.
        \item For every $ 1 \le h \le k - 1 $ it holds $ \mathtt{t}_{u'_h} = \mathtt{t}_{v_h} = 3 $. In this case, for every $ 1 \le h \le k - 1 $ we have $ \mathtt{t}_{v_h} = 3 $ and $ (\bar{v}_{h + 1}, \bar{v}_h) \in \bar{E} $, so $ v_{h + 1} \in G_{\ell_{v_h}}(v_h) $ and by Lemma \ref{lem:propertiesexploringminimum} there exists an occurrence of $ \min_{v_h}[1, \ell_{v_h} - 1] = \gamma_{v_h}[1, \ell_{v_h} - 1] = \gamma_{u'_h}[1, \ell_{u'_h} - 1] = \min_{u'_h}[1, \ell_{u'_h} - 1] $ starting at $ v_h $ and ending at $ v_{h + 1} $. As a consequence, there is an occurrence of $ \min_{u'_1}[1, \ell_{u'_1} - 1] \min_{u'_2}[1, \ell_{u'_2} - 1] \dots \min_{u'_{k - 1}}[1, \ell_{u'_{k - 1}} - 1] = \min_u[1, j_k - 1] $ starting at $ u $ and ending at $ v_k $ and so, if $ \beta' = \min_u[1, j_k - 1] \min_{v_k} $, then $ \beta' \in I_u $. This implies $ \min_u \preceq \beta' $, so from $ \min_u = \min_u[1, j_k - 1] \min_{u'_k} $ we obtain $ \min_{u'_k} \preceq \min_{v_k} $.
        By Lemma \ref{lem_motivatingorder} and $ \min_{u'_k} \preceq \min_{v_k} $, we obtain that one of the following statements must be true:
        \begin{enumerate}
            \item $ \gamma_{u'_k} $ is not a prefix of $ \gamma_{v_k} $, $ \gamma_{v_k} $ is not a prefix of $ \gamma_{u'_k} $ and $ \gamma_{u'_k} \prec \gamma_{v_k} $.
            \item $ \gamma_{u'_k} = \gamma_{v_k} $ and $ \mathtt{t}_{u'_k} \le \mathtt{t}_{v_k} $.
            \item $ \gamma_{v_k} $ is a strict prefix of $ \gamma_{u'_k} $.
        \end{enumerate}
        In all three cases, we conclude $ (\gamma_{u'_k}, \mathtt{t}_{u'_k}) \preceq' (\gamma_{v_k}, \mathtt{t}_{v_k}) $, or equivalently, $ \lambda (u'_k) \preceq' \lambda (v_k) $, or equivalently, $ \alpha[k] \preceq' \beta[k] $.
    \end{enumerate}
\end{proof}

\noindent{\textbf{Statement of Theorem \ref{theor:reductiontosmaller}}.}
    Let $ G = (V, E) $ be a graph. Let $ u, v \in V $ be such that $ \tau(u) = \tau(v) = 3 $.
    \begin{enumerate}
        \item If $ \min_u = \min_v $, then $ \min_{\bar{u}} = \min_{\bar{v}} $.
        \item If $ \min_u \prec \min_v $, then $ \min_{\bar{u}} \prec' \min_{\bar{v}} $.
    \end{enumerate}

\begin{proof}
    Let $ (u_i)_{i \ge 1} $ be and occurrence of $ \min_u $ starting at $ u $, and let $ (v_i)_{i \ge 1} $ be and occurrence of $ \min_v $ starting at $ v $ . Let $ (\bar{u}'_i)_{i \ge 1} $ be the occurrence of $\min_{\bar{u}} $ defined by means of $ (u_i)_{i \ge 1} $ in Lemma \ref{ref:minimumquotientgraph}, and let $ (\bar{v}'_i)_{i \ge 1} $ be the occurrence of $\min_{\bar{v}} $ defined by means of $ (v_i)_{i \ge 1} $ in Lemma \ref{ref:minimumquotientgraph}. Let $ L = \{i \ge 1 \;|\; \tau(u_i) = 3 \} $, and let $ j_i \ge 1 $ be the $ i^{\text{th}} $ smallest element of $ L $, if it exists. Moreover, let $ M = \{i \ge 1 \;|\; \tau(v_i) = 3 \} $, and let $ k_i \ge 1 $ be the $ i^{\text{th}} $ smallest element of $ K $, if it exists. Notice that $ j_1 = k_1 = 1 $. 
    
    In the rest of the proof, we say that an integer $ h \ge 1 $ is \emph{nice} if it satisfies the following properties:
        \begin{itemize}
            \item $ j_1, \dots, j_h $ and $ k_1, \dots, k_h $ are all defined, and $ j_i = k_i $ for every $ 1 \le i \le h $.
            \item $ \gamma_{u'_i} = \gamma_{v'_i} $ for every $ 1 \le i \le h - 1 $.
        \end{itemize}
        Note the following properties:
        \begin{itemize}
            \item $ 1 $ is always nice because $ j_1 = k_1 = 1 $.
            \item If $ h $ is nice, than every $ 1 \le h' \le h $ is nice. This implies that either every $ h \ge 1 $ is nice, or there exists  a unique $ h^* \ge 1 $ such that every $ h \le h^* $ is nice and every $ h > h^* $ is not nice.
            \item If $ h $ is nice, $ \gamma_{u'_h} = \gamma_{v'_h} $ and $ \mathtt{t}_{u'_h} = \mathtt{t}_{v'_h} = 3 $, then $ h + 1 $ is nice. Indeed, $ \gamma_{u'_h} = \gamma_{v'_h} $ implies $ \ell_{u'_h} = \ell_{v'_h} $. Moreover, since $ (u_i)_{i \ge j_h} $ is an occurrence of $ \min_{u'_h} $ starting at $ u'_h $, then by the minimality of $ \ell_{u'_h} $ and Lemma \ref{lem:propertiesexploringminimum} we have $ \tau(u_{j_h + i}) = \tau(G_{i + 1}(u'_h)) = 1 $ for every $ 1 \le i \le \ell_{u'_h} - 2 $, and $ \tau(u_{j_h + \ell_{u'_h} - 1}) = \tau(G_{\ell_{u'_h}}(u'_h)) = \mathtt{t}_{u'_h} = 3 $, which implies that $ j_{h + 1} $ is defined. Analogously, one obtains $ \tau(u_{k_h + i}) = 1 $ for every $ 1 \le i \le \ell_{v'_h} - 2 $ and $ \tau(v_{k_h + \ell_{v'_h} - 1}) = 3 $, so $ k_{h + 1} $ is defined and $ j_{h + 1} = k_{h + 1} $, which implies that $ h + 1 $ is nice.
            \item If $ h $ is nice, then $ \mathtt{t}_{u'_i} = \mathtt{t}_{v'_i} = 3 $ for every $ 1 \le i \le h - 1 $. Indeed, assume for sake of contradiction that $ \mathtt{t}_{u'_l} = 2 $ for some $ 1 \le l \le h - 1 $ (the case $ \mathtt{t}_{v'_l} = 2 $ is analogous). Since $ (u_i)_{i \ge j_l} $ is an occurrence of $ \min_{u'_l} $ starting at $ u'_l $, then by Lemma \ref{lem:propertiesexploringminimum} we have $ \tau(u_{j_l + i}) = \tau(G_{i + 1}(u'_l)) = 1 $ for every $ 1 \le i \le \ell_{u'_l} - 2 $, and $ \tau(u_{j_l + \ell_{u'_l} - 1}) = \tau(G_{\ell_{u'_l}}(u'_l)) = \mathtt{t}_{u'_l} = 2 $, which by Lemma \ref{lem:descrptionminimum} implies $ \tau(u_i) = 2 $ for every $ i \ge j_l + \ell_{u'_l} - 1 $. This implies that $ j_{l + 1} $ is not defined, which is a contradiction because $ 1 \le l \le h - 1 $.
        \end{itemize}
        Let us prove that, if $ h \ge 1 $ is nice, then:
        \begin{itemize}
            \item $ \min_u[1, j_h - 1] = \min_v[1, k_h - 1] $.
            \item $ \min_{\bar{u}}[1, h -1] = \min_{\bar{v}}[1, h - 1] $.
        \end{itemize}
        We will prove these two properties separately.
        \begin{itemize}
            \item Let us prove that $ \min_u[1, j_h - 1] = \min_v[1, j_h - 1] $. Since $ h $ is nice, then for every $ 1 \le i \le h $  we have that $ j_{i} $ and $ k_{i} $ are defined, and $ j_{i} = k_{i} $. As a consequence, it will suffice to prove that $ \min_u[j_{i}, j_{i + 1} - 1] = \min_v[k_{i}, k_{i + 1} - 1] $ for every $ 1 \le i \le h - 1 $. This is equivalent to proving that $ \min_{u'_i}[1, \ell_{u'_i} - 1] = \min_{v'_i}[1, \ell_{v'_i} - 1] $ for every $ 1 \le i \le h - 1 $. This follows from $ \gamma_{u'_i} = \gamma_{v'_i} $ for every $ 1 \le i \le h - 1 $.
            \item Let us prove that $ \min_{\bar{u}}[1, h -1] = \min_{\bar{v}}[1, h - 1] $. Fix $ 1 \le i \le h - 1 $. We must prove that $ \min_{\bar{u}}[i] = \min_{\bar{v}}[i] $, or equivalently, $ \lambda(\bar{u}'_i) = \lambda(\bar{v}'_i) $. This follows from $ \gamma(u'_i) = \gamma(v'_i) $ and $ \mathtt{t}_{u'_i} = \mathtt{t}_{v'_i} $.
        \end{itemize}
        
        We can now prove the two claims of the theorem. We will use that following observation: if every $ h \ge 1 $ is nice, then $ \min_u = \min_v $ and $ \min_{\bar{u}} = \min_{\bar{v}} $, because $ \min_u[1, j_h - 1] = \min_v[1, k_h - 1] $ and $ \min_{\bar{u}}[1, h -1] = \min_{\bar{v}}[1, h - 1] $ for every $ h \ge 1 $.
        \begin{enumerate}
            \item Assume that $ \min_u = \min_v $; we must prove that $ \min_{\bar{u}} = \min_{\bar{v}} $. If every $ h \ge 1 $ is nice, we are done, so we can assume that $ h^* \ge 1 $ is the largest nice integer. By Lemma \ref{lem:comparingminima}, we have $ \tau (u_i) = \tau (v_i) $ for every $ i \ge 1 $, so $ L = M $. In particular, for every $ i \ge 1 $ we have that $ j_i $ is defined if and only if $ k _i $ is defined and, if so $ j_i = k_i $. Since $ h^* $ is nice, then $ j_{h^*} $ and $ k_{h^*} $ are defined, $ j_{h^*} = k_{h^*} $ and $ \min_{\bar{u}}[1, h^* -1] = \min_{\bar{v}}[1, h^* - 1] $. We know that $ \min_{\bar{u}} = \min_{\bar{u}}[1, h^* -1] \min_{\bar{u}'_{h^*}} $ and $ \min_{\bar{v}} = \min_{\bar{v}}[1, h^* -1] \min_{\bar{v}'_{h^*}} $, so $ \min_{\bar{u}}[1, h^* -1] = \min_{\bar{v}}[1, h^* - 1] $ implies that in order to prove that $ \min_{\bar{u}} = \min_{\bar{v}} $ we only have to prove that $ \min_{\bar{u}'_{h^*}} = \min_{\bar{v}'_{h^*}} $.

            By Lemma \ref{lem:comparingminima}, we have $ \min_{u'_{h^*}} = \min_{v'_{h^*}} $. Since $ \tau (u_i) = \tau (v_i) $ for every $ i \ge 1 $, $ (u_i)_{i \ge j_{h^*}} $ is an occurrence of $ \min_{u'_{h^*}} $ starting at $ u'_{h^*} $ and $ (v_i)_{i \ge j_{h^*}} $ is an occurrence of $ \min_{v'_{h^*}} $ starting at $ v'_{h^*} $, we obtain $ \ell_{u'_{h^*}} = \ell_{v'_{h^*}} $. As a consequence, $ \gamma_{u'_{h^*}} = \min_{u'_{h^*}}[1, \ell_{u'_{h^*}}] = \min_{v'_{h^*}}[1, \ell_{v'_{h^*}}] = \gamma_{v'_{h^*}} $. In particular, it must be $ \mathtt{t}_{u'_{h^*}} = \mathtt{t}_{v'_{h^*}} $, so from $ \gamma_{u'_{h^*}} = \gamma (v'_{h^*}) $ we conclude $ \lambda (\bar{u}'_{h^*}) = \lambda (\bar{v}'_{h^*}) $. Moreover, it must be $ \mathtt{t}_{u'_{h^*}} = \mathtt{t}_{v'_{h^*}} = 2 $, otherwise we would conclude that $ h^* + 1 $ is nice, which contradicts the maximality of $ h^* $. As a consequence, by the definition of $ \bar{G} = (\bar{V}, \bar{E}) $ we conclude that $ \min_{\bar{u}'_{h^*}} = (\lambda (\bar{u}'_{h^*}))^\omega = \lambda (\bar{v}'_{h^*}))^\omega = \min_{\bar{v}'_{h^*}} $.
            \item Assume that $ \min_u \prec \min_v $; we must prove that $ \min_{\bar{u}} \prec' \min_{\bar{v}} $. Notice that it cannot happen that every $ h \ge 1 $ is nice, otherwise we would obtain $ \min_u = \min_v $, a contradiction. Let $ h^* \ge 1 $ be the biggest nice integer. In particular, $ j_{h^*} $ and $ k_{h^*} $ are defined, $ j_{h^*} = k_{h^*} $, $ \min_u[1, j_{h^*} - 1] = \min_v[1, k_{h^*} - 1] $ and $ \min_{\bar{u}}[1, h^* -1] = \min_{\bar{v}}[1, h^* - 1] $. We know that $ \min_u = \min_u[1, j_{h^*} - 1] \min_{u'_{h^*}} $ and $ \min_v = \min_v[1, k_{h^*} - 1] \min_{v'_{h^*}} $, so from $ \min_u \prec \min_v $ we conclude $ \min_{u'_{h^*}} \prec \min_{v'_{h^*}} $. Since $ \min_{\bar{u}} = \min_{\bar{u}}[1, h^* - 1] \min_{\bar{u}'_{h^*}} $, $ \min_{\bar{v}} = \min_{\bar{v}}[1, h^* - 1] \min_{\bar{v}'_{h^*}} $ and $ \min_{\bar{u}}[1, h^* -1] = \min_{\bar{v}}[1, h^* - 1] $, in order to prove that $ \min_{\bar{u}} \prec' \min_{\bar{v}} $ we only have to prove that $ \min_{\bar{u}'_{h^*}} \prec' \min_{\bar{v}'_{h^*}} $.
            
            Notice that it cannot be $ \gamma_{u'_{h^*}} = \gamma_{v'_{h^*}} $ and $ \tau(u'_{h^*}) = \tau(v'_{h^*}) $, because:
        \begin{enumerate}
            \item If $ \tau(u'_{h^*}) = \tau(v'_{h^*}) = 2 $, then by Lemma \ref{lem:type2equalityminima} we would conclude $ \min_{u'_{h^*}} = \min_{v'_{h^*}} $, a contradiction.
            \item If $ \tau(u'_{h^*}) = \tau(v'_{h^*}) = 3 $, then $ h^* + 1 $ would be a nice integer, which contradicts the maximality of $ h^* $.
        \end{enumerate}
        From this observation, Lemma \ref{lem_motivatingorder} and $ \min_{u'_{h^*}} \prec \min_{v'_{h^*}} $ we conclude that one of the following must be true:
        \begin{enumerate}
            \item $ \gamma_{u'_{h^*}} $ is not a prefix of $ \gamma_{v'_{h^*}} $, $ \gamma_{v'_{h^*}} $ is not a prefix of $ \gamma_{u'_{h^*}} $ and $ \gamma_{u'_{h^*}} \prec \gamma_{v'_{h^*}} $.
            \item $ \gamma_{u'_{h^*}} = \gamma_{v'_{h^*}} $, $ \mathtt{t}_{u'_{h^*}} = 2 $ and $ \mathtt{t}_{v'_{h^*}} = 3 $.
            \item $ \gamma_{u'_{h^*}} $ is a strict prefix of $ \gamma_{v'_{h^*}} $.
        \end{enumerate}
        In all three cases we conclude $ (\gamma_{u'_{h^*}}, \mathtt{t}_{u'_{h^*}}) \prec' (\gamma_{v'_{h^*}}, \mathtt{t}_{v'_{h^*}}) $, or equivalently, $ \lambda(\bar{u}'_{h^*}) \prec' \lambda(\bar{v}'_{h^*}) $, which implies $ \min_{\bar{u}'_{h^*}} \prec' \min_{\bar{v}'_{h^*}} $.
        \end{enumerate}
 
\end{proof}

\noindent{\textbf{Statement of Lemma \ref{lem:runningtimderivedgraph}}.}
    Let $ G = (V, E) $ be a trimmed graph. Then, we can build $ \bar{G} = (\bar{V}, \bar{E}) $ in $ O(|V|^2) $ time.

\begin{proof}
    The definition of $ \bar{G} $ implies that in order to build $ \bar{G} $ it is sufficient to compute $ \gamma_u $, $ \mathtt{t}_u $ and $ G_{\ell_u} (u) $ for every $ u \in V $ such that $ \tau (u) = 3 $; moreover, we also need to compute the total order $ \preceq' $, so that we can remap $ \Sigma' $ to an integer alphabet consistently with the total order $ \preceq' $ on $ \Sigma' $.

    Fix $ u \in V $ such that $ \tau (u) = 3 $. For every $ 1 \le i \le \ell_u - 1 $, let $ E_i(u) $ be the set of all edges entering a node in $ G_i (u) $, and let $ m_i (u) = |E_i(i)| $. 
    Note that since the $ G_i (u) $'s are pairwise distinct, then the $ E_u (i) $'s are pairwise disjoint.
    Let us prove that $ \sum_{i = 1}^{\ell_u - 1} m_i (u) \in O(n) $. To this end, it will suffice to prove that for every $ v \in V $ there exist at most two edges in $ \bigcup_{i = 1}^{\ell_u - 1} E_i(u) $ whose start nodes are equal to $ v $. Fix $ v \in V $, and let $ 1 \le i \le \ell_u - 1 $ be the smallest integer such that $ v $ is the first state of an edge $ (v, v') \in E_i(u) $ for some $ v' \in G_i (u) $, if such a $ v' $ exists. Notice that $ v' $ is uniquely determined because the value $ \lambda (z) $ does not depend on the choice of $ z \in G_i (u) $, and $ G $ is deterministic. This means that $ (v, v') \in E_i(u) $ is uniquely determined. In order to prove our claim, it will suffice to show that if $ (v, v'') \in E_j(u) $ for some $ i + 1 \le j \le \ell_u - 1 $ and  $ v'' \in G_j (u) $, then $ \lambda (v'') = \lambda (v) $, because we will conclude that $ j $ and $ v'' $ are uniquely determined, since $ G $ is deterministic. Since $ v' \in G_i (u) $ and $ (v, v') \in E_i (u) $, by the definition of $ G_{i + 1} (u) $, we have $ \lambda (G_{i + 1} (u)) \preceq \lambda (v) $. Since $ 2 \le i + 1 \le j \le \ell_u - 1 $, then by Lemma \ref{lem:charatersdecrease} we obtain $ \min_u [j] \preceq \min_u [i + 1] $, which by Lemma \ref{lem:propertiesexploringminimum} is equivalent to $ \lambda (G_{j}(u)) \preceq \lambda (G_{i + 1} (u)) $, and so we conclude $ \lambda (G_{j} (u)) \preceq \lambda (v) $, or equivalently, $ \lambda (v'') \preceq \lambda (v) $. Since $ v'' \in G_{j} (u) $ and $ 2 \le j \le \ell_u - 1 $, then $ \tau (v'') = \tau (G_{j} (u)) = 1 $, so from $ (v, v'') \in E $ we conclude that it must be $ \lambda (v) \preceq \lambda (v'') $ because $ G $ is trimmed, and so $ \lambda (v'') = \lambda (v) $.

    Let us show that in $ O(|V|^2) $ we can compute  $ G_{\ell_u}(u) $, $ \gamma_u $ and $ \mathtt{t}_u $ for every $ u \in V $ such that $ \tau (u) = 3 $. It will suffice to show that, for a a fixed $ u \in V $ such that $ \tau (u) = 3 $, we can compute $ G_{\ell_u}(u) $, $ \gamma_u $ and $ \mathtt{t}_u $ in $ O(|V|) $ time.

    Let us show how we recursively compute each set $ G_i (u) $, for $ 1 \le i \le \ell_u $. During the algorithm, we will mark some nodes. Initially, no node is marked, and after step $ i \ge 1  $, a nodes of $ V $ is marked if and only if it belongs to $ \bigcup_{j = 2}^{i - 1} G_j (u) $. If $ i = 1 $, we just let $ G_1 (u) = \{u \} $, and we define $ c_1 = \lambda (u) $. Now, assume that $ i \ge 2 $, and assume that we have computed $ G_{i - 1}(u) $. We first scan all edges in $ E_{i - 1}(u) $  and we collect the set  $ R_{i} $ the set of all start nodes of these edges. Then, by scanning $ R_{i} $, we first determine $ c_{i} = \min\{\lambda (v) \;|\; v \in R_{i} \} $, then we determine $ R'_{i} = \{v \in R_{i} \;|\; \lambda (v) = c_{i} \} $. Next, we determine $ t_{i} = \min\{\tau(v) \;|\; v \in R'_{i} \} $ and $ R''_{i} = \{v \in R'_{i} \;|\; \lambda (v) = t_{i} \} $. By checking which nodes in $ R''_{i} $ are marked, we determine $ G_i (u) $. Finally, we mark all nodes in $ G_i (u) $. We can perform all these steps in $ O(m_{i - 1} (u)) $ time, where the hidden constant in the asymptotic notation does not depend on $ i $. Now, we check whether $ t_{i} \ge 2 $. If $ t_{i} \ge 2 $, then $ i = \ell_u $ and we are done. Otherwise, we have $ i < \ell_u $ and we proceed with step $ i + 1 $.

    We conclude that we can determine $ G_{\ell_u}(u) $ in time $ \sum_{i = 1}^{\ell_u - 1} O(m_i (u)) = O(\sum_{i = 1}^{\ell_u - 1} m_i (u)) = O(n) $, where we have used that the hidden constant in $ O(m_i (u)) $ does not depend on $ i $. In addition, we have $ \gamma_u = c_1 c_2 \dots c_{\ell_u} $ and $ \mathtt{t}_u = t_{\ell_u} $.

   We are only left with showing how to compute $ \preceq' $. Essentially, we only have to radix sort the strings $ \gamma (u) $'s by taking into account that $ \preceq' $ is defined by considering a slight variation of the lexicographic order. More precisely, we proceed as follows. We know that $ |\gamma (u)| \le |V| + 1 $ for each $ u \in V $ such that $ \tau (u) = 3 $, so we first pad the end of each $ \gamma (u) $ with a special character larger than all characters in the alphabet until the length of each string is exactly $ |V| + 1 $. Next, we consider two extra special characters $ d_2 $ and $ d_3 $ such that $ d_2 \prec' d_3 $, and we append exactly one of this character to each $ \gamma (u) $: we append $ d_2 $ if $ \mathtt{t}_u = 2 $, and we append $ d_3 $ if $ \mathtt{t}_u = 3 $. Now, we radix sort the (modified) $ \gamma (u) $'s in $ O(|V|^2) $ time, so obtaining $ \preceq' $.
\end{proof}

\subsection{Proofs from Section \ref{sec:merging}} 

\noindent{\textbf{Statement of Lemma \ref{lem:detailsmerging}}.}
    Let $ G = (V, E) $ be a graph. If we know the min-partition of $ \{u \in V \;|\; \tau (u) = 3 \} $, then we can build the min-partition of $ V $ in $ O(|E|) $ time. 

\begin{proof}
    We have seen that the algorithm correctly builds $ \mathcal{A}_{c, 3} $ and $ \mathcal{A}_{c, 2} $ for every $ c \in \Sigma $. Note that when the algorithm builds the $ \mathcal{A}_{c, 1} $'s, it never modifies the $ A_{c, 2} $'s and the $ A_{c, 3} $'s (because we only add elements to the $ A_{c, 1} $'s).



    Let us prove that, when we consider $ I $ in $A_{c_i, t} $ (and before computing the $ J_k $'s), then:
    \begin{enumerate}
        \item $ I $ is an element of $ \mathcal{A} $ and we have already built the prefix of $ \mathcal{A} $ whose largest element is $ I $.
        \item every $ A_{c, 1} $ contains a prefix of $ \mathcal{A}_{c, 1} $.
    \end{enumerate}

    At the beginning of the algorithm we consider $ I $ in $ A_{c_1, t} $, with $ t \in \{2, 3 \} $, so our claim is true because we have already built $ \mathcal{A}_{c_1, 2} $ and $ \mathcal{A}_{c_1, 3} $, and all the $ A_{c, 1} $'s are empty. 
    
    Now, assume that our claim is true when we consider $ I \in \mathcal{A}_{c_i, t} $. We want to prove that it is true when we process the next element (if it exists). When we consider $ I $, we compute the nonempty $ J_k $'s, and we add each such $ J_k $ to $ A_{c_k, 1} $. We now want to prove that $ J_k $ is correctly identified as the next element in $ A_{c_k, 1} $.
    \begin{itemize}
        \item First, let us prove that, if $ v_1, v_2 \in J_k $, then $ \min_{v_1} = \min_{v_2}$. Since $ v_1, v_2 \in J_k $, then there exist $ u_1, u_2 \in I $ such that $ (u_1, v_1), (u_2, v_2) \in E $. Since by the inductive hypothesis $ I $ is an element of $ \mathcal{A} $ and we have already built the prefix of $ \mathcal{A} $ whose largest element is $ I $ and since $ v_1 $ and $ v_2 $ were not marked before, then it must be $ \min_{v_1} = c_k \min_{u_1} $, $ \min_{v_2} = c_k \min_{u_2}  $ and $ \min_{u_1} = \min_{u_2} $, so we conclude $ \min_{v_1} = \min_{v_2} $.
        \item We are only left with showing that if $ v_1 \in J_k $ and $ v_2 \in V_{c_k, 1} $ is not in some element of $A_{c_k, 1} $ after $ J_k $ is added to $ A_{c_k, 1} $, then $ \min_{v_1} \prec \min_{v_2} $. Let $ u_1 \in I $ such that $ (u_1, u_2) \in E $; we have shown that $ \min_{v_1} = c_k \min_{u_1} $. The definition of $ v_2 $ implies that $ \min_{v_2} = c_k \min_{u_2} $ for some $ u_2 $ that we have not processed yet. Since by the inductive hypothesis $ I $ is an element of $ \mathcal{A} $ and we have already built the prefix of $ \mathcal{A} $ whose largest element is $ I $, then it must be $ \min_{u_1} \prec \min_{u_2} $, and we conclude $ \min_{v_1} \prec \min_{v_2} $.
    \end{itemize}
    Now, let us prove that, if after considering $ I $ in $ A_{c_i, t} $, the next element to be considered is not in $ A_{c_i, t} $, then the construction of $ \mathcal{A}_{c_1, t} $ is complete. If $ t \in \{2, 3 \} $ then the conclusion is immediate because $ \mathcal{A}_{c_i, 2} $ and $ \mathcal{A}_{c_i, 3} $ had already been fully built. Now assume that $ t = 1 $. Suppose for sake of contradiction that there exists $ v \in V_{c_i, 1} $ which has not been added to some element in $ A_{c_i, 1} $. Without loss of generality, we choose $ v $ such that $ \min_v $ is as small as possible. Since $ \tau (v) = 1 $, then there exists $ u \in V $ such that $ (u, v) \in E $ and $ \min_u \prec \min_v $ (and so $ \lambda (u) \preceq \lambda (v) = c_i $). If $ \lambda (u) \prec c_i $, then we immediately obtain that $ u $ has already been processed (because we have already built the prefix of $ \mathcal{A} $ whose largest element is $ I $) and so $ v $ should have already been processed when $ u $ was processed, a contradiction. If $ \lambda (u) = c_i $, then by Corollary \ref{cor:typesproperties} we have $ \tau (u) \le \tau (v) $, so $ \tau (u) = 1 $ and $ u \in V_{c_i, 1} $; the minimality of $ v $ implies that $ u $ was previously added to some element of $ A_{c_i, 1} $ and so $ v $ should have been processed, again a contradiction.

    As a consequence, when we process the next element after $ I $ in $ A_{c_i, t} $, then our claim will be true (independently of whether the next element is in $ A_{c_i, t} $ or not). In particular, if there is no next element, we conclude that all $ \mathcal{A}_{c, 1} $'s have been built,

    Lastly, we can build each $ \mathcal{A}_{c, 1} $ in $ O(|E|) $ time, because we only need to scan each node and each edge once.

\end{proof}

\section{Details on the Complementary Case}

Let us show how to reduce the problem of computing the min-partition of $ \{u \in V \;|\; \tau (u) = 1 \} $ to the problem of determining the min-partition of $ \bar{V} $, where $ \bar{G} = (\bar{V}, \bar{E}) $ is a graph such that $ |\bar{V}| = \{u \in V \;|\; \tau (u) = 1 \} $. We use the same notation that we used in the previous sections (with a complementary meaning) so that it will easy to follow our argument.

\subsection{Recursive Step}

For every $ u \in V $ such that $ \tau (u) = 1 $, let $ \ell_u $ be the smallest integer $ k \ge 2 $ such that $ \tau(u_k) \le 2 $, where $ (u_i)_{i \ge 1} $ is an occurrence of $ \min_u $ starting at $ u $. For every $ 1 \le i \le \ell_u $, we can define each $ G_i (u) $ as we did before; now, it will be $ \tau (G_i (u)) = 3 $ for every $ 2 \le i \le \ell_u - 1 $. In fact, when we explore the graph starting from $ u $ in a backward fashion and we (implicitly) build a prefix of $ \min_u $, at each step it is still true that we must consider the nodes with minimum value $ \lambda (v) $ and, among those, the nodes with minimum value $ \tau(v) $. This time, there is no chance that we include in $ G_i (u) $ a node already included before: at every step, $ \lambda (G_i (u)) $ can only increase (because $ \tau (G_i (u)) = 3 $), so if there were a cycle, then all edges of the cycles would have the same label and by Lemma \ref{lem:characterizationtype2} we would conclude $ \tau (G_i (u)) \le 2 $ for some $ 2 \le i \le \ell_u - 1 $, a contradiction. In particular, this means that now we do not even need to somehow trim the graph to ensure that our algorithm runs in $ O(|V|^2) $ time, because we cannot visit the same node twice. The $ \gamma_u $'s and the $ \mathtt{t}_u $'s are defined as before, and Lemma \ref{lem_motivatingorder} still implies that $ \Sigma' $ and $ \preceq' $ must be defined as before. When we define $ \bar{G} = (\bar{V}, \bar{E}) $, we define $ \bar{V} = \{\bar{u} \;|\; \tau (u) = 1 \} $ and we define $ \bar{E} $ as before. It is easy to check that we can build $ \bar{G} $ in $ O(|V|^2) $ time as before, and if we have the min-partition of $ \bar{V} $ (with respect to $ \bar{G} $), then we also have the min-partition of $ \{u \in V \;|\; \tau (u) = 1 \} $ (with respect to $ G $).

\subsection{Merging}

Let $ u \in V $ such that $ \tau (u) = 3 $. By Lemma \ref{lem:typeequivalent}, there exist $ k \ge 1 $, $ c \in \Sigma $ and $ \gamma' \in \Sigma^\omega $ such that $ \min_u = \lambda (u)^k c \gamma' $ and $ \lambda (u) \prec c $. Then, we define $ \psi (u) = k $.

In the following, we will often use the following observation. Let $ v \in V $ such that $ \tau (v) = 3 $. Let $ u \in V $ such that $ (u, v) \in E $ and $ \min_v = \lambda (v) \min_u $. Then, (i) $ \lambda (v) \preceq \lambda (u) $ and (ii) if $ \lambda (v) = \lambda (u) $, then $ \tau (u) = 3 $ and $ \psi (v) = \psi (u) + 1 $.

Note that, if $ u, v \in V $ are such that $ \tau (u) = \tau (v) = 3 $, $ \lambda (u) = \lambda (v) $ and $ \psi (u) \not = \psi (v) $, then $ \min_u \prec \min_v $ if and only if $ \psi (v) < \psi (u) $.

It is easy to compute all $ \psi (u) $'s in $ O(|E|) $ time. Start from each $ u \in V $ such that $ \tau (u) = 3 $, and explore the graph in a backward fashion by only following edges labeled $ \lambda (u) $, until either we encounter a node for which we have already determined $ \psi (u) $, or we cannot explore the graph any longer. For all the nodes $ u' $ that we encounter it must be $ \tau (u') = 3 $ (otherwise it would be $ \tau (u) \not = 3 $), and if we cannot explore any longer from some $ u' $ that we encounter, it must be $ \psi (u') = 1 $. We cannot encounter the same node twice because $ G $ is deterministic, and we cannot have cycles otherwise it would be $ \tau (u) \not = 3 $ by Lemma \ref{lem:characterizationtype2}. As a consequence, we have built a tree (where the root $ u $ has no \emph{outgoing} edges), and computing the $ \psi (u) $'s is equivalent to computing the height of each node.

Let us show how we can use the $ \psi (u) $'s to determine the min-partition $ \mathcal{A} $ of $ V $, assuming that we already have the min-partition $ \mathcal{B}$ of $ \{u \in V \;|\; \tau (u) = 1 \} $.
As before we have the min-partition $ \mathcal{B}' $ of $ \{ u \in \mathcal{V} \;|\; \tau (u) = 2 \} $. For every $ c \in \Sigma $ and for every $ t \in \{1, 2, 3 \} $, we define $ V_{c, t} $, $ \mathcal{A}_{c, t} $ and $ A_{c, t} $ as before, and our problem reduces to compute each $ \mathcal{A}_{c, t} $.

By using $ \mathcal{B} $ and $ \mathcal{B}' $, we can compute the $ \mathcal{A}_{c, 1} $'s and the $ \mathcal{A}_{c, 2} $'s as before, so the challenging part is to compute the $ \mathcal{A}_{c, 3} $'s. Let $ \Sigma = \{c_1, c_2, \dots, c_{\sigma} \} $, with $ c_1 \prec c_2 \prec \dots \prec c_\sigma $. During the algorithm, we will assign a number $ \psi(I) \ge 1 $ to every element $ I $ being in some $ A_{c, 3} $ at some point.

Notice that $ \mathcal{A}_{c_\sigma, 3} $ must be empty because otherwise we would conclude that $ c_{\sigma} $ is not largest character. This suggest that this time the order in which we process the $ \mathcal{A}_{c, t} $'s must be $ \mathcal{A}_{c_\sigma, 2} $, $ \mathcal{A}_{c_\sigma, 1} $, $ \mathcal{A}_{c_{\sigma - 1}, 3} $, $ \mathcal{A}_{c_{\sigma - 1}, 2} $, $ \mathcal{A}_{c_{\sigma - 1}, 1} $, $ \mathcal{A}_{c_{\sigma - 2}, 3} $ and so on. Moreover, we will build each $ \mathcal{A}_{c, t} $ incrementally from its largest element to its smallest element (so we will consider \emph{suffixes} of min-partitions, not prefixes). This time we will not mark nodes, but we will mark entries of the $ A_{c, t} $'s to indicate that a node has been removed from an element in $ A_{c, t} $. Intuitively, this time we need to remove nodes because it will be true that when we process $ I $ in $ A_{c_i, t} $ then we have already built the suffix (not the prefix) of $ \mathcal{A} $ whose \emph{largest} element is $ I $, but we are now building $ \mathcal{A} $ from its largest element to its smallest element, so a node reached by an edge leaving a node $ u $ in $ I $ may also be reached by an edge leaving a node $ u' $ that we have not processed, and for which it holds $ \min_{u'} \prec \min_u $.

Assume that we process $ I $ in $ A_{c_i, t} $. Note that if $ v \in V $ is such that $ \tau (v) = 3 $, $ \lambda (v) = c_k $, and $ (u, v) \in E $ for some $ u \in I $, then it must be $ k \le i $ otherwise $ \tau (v) = 1  $; if $ I $ is an element in the $ \mathcal{A}_{c, 1} $'s or in the $ \mathcal{A}_{c, 2} $'s, it must always be $ k < i $ because, if it were $ k = i $, then we would conclude $ \tau (v) \not = 3 $. For every $ k \le i $, define $ \psi_{I, k} = 1 $ if $ k < i $, and $ \psi_{I, k} = \psi(I) + 1 $ if $ k = i $. We compute the set $ J_k $ of all nodes $ v \in V $ such that $ \tau (v) = 3 $, $ \lambda (v) = c_k $, $ \psi (v) = \psi_{I, k} $ and $ (u, v) \in E $ for some $ u \in I $ and, if $ J_k \not = \emptyset $, then (i) we mark the entries of $ A_{c_k, 3} $ containing an element in $ J_k $ and (ii) we add $ J_k $ to $ A_{c_k, 3} $, letting $ \psi(J_k) = \psi_{I, k} $. We assume that we maintain an additional array that for every node in $ \{ u \in \mathcal{V} \;|\; \tau (u) = 3 \} $ already occurring in some $ A_{c, 3} $ stores its (unique) current position, so that operation (i) can be performed in constant time without affecting the running time of the algorithm (which is still $ O(|E|) $).

Notice that at any time the following will be true in each $ A_{c, 3} $: (a) if $ K $ is in $ A_{c, 3} $, then $ \psi(K) \ge 1 $; (b) if $ A_{c, 3} $ is nonempty, then the first $ K $ that we have added is such that $ \psi(K) = 1 $; (c) If $ K' $ has been added to $ A_{c, 3} $ immediately after $ K $, then $ \psi(K) \le \psi(K') \le \psi(K) + 1 $; (d) If $ v \in K $ for some $ K $ in $ A_{c, 3} $, then $ \psi (v) = \psi (K) $.

Let us prove that, when we consider $ I $ in $A_{c_i, t} $ (and before computing the $ J_k $'s), then $ I $ is an element of $ \mathcal{A} $ and we have already built the suffix of $ \mathcal{A} $ whose largest element is $ I $.

At the beginning of the algorithm we consider $ I $ in $ A_{c_\sigma, t} $, with $ t \in \{1, 2 \} $, so our claim is true because we have already built $ \mathcal{A}_{c_\sigma, 2} $ and $ \mathcal{A}_{c_\sigma, 1} $.

Now, assume that our claim is true when we consider $ I $ in $ A_{c_i, t} $. Let us prove that we can consider the next element $ I' $ in $ A_{c_i', t'} $ (if it exists), then $ I' $ is an element of $ \mathcal{A}$ and we have already built the suffix of $ \mathcal{A} $ whose largest element is $ I' $. Notice that either $ i' = i $ or $ i' = i - 1 $.
\begin{enumerate}
    \item Assume that $ i' = i $ and $ t' = t $. If $ t' = t \not = 3 $ we are done because we have already built the $ \mathcal{A}_{c, 2} $'s and the $ \mathcal{A}_{c, 1} $'s. Now assume that $ t' = t = 3 $. This implies that $ \psi (I) + 1 \le \psi (I') \le \psi (I) $.
    \begin{enumerate}
        \item Suppose that $ \psi (I') = \psi (I) $. First, let us prove that, if $ v_1, v_2 \in I' $, then $ \min_{v_1} = \min_{v_2} $. Let $ u_1, u_2 \in V $ be such that $ (u_1, v_1), (u_2, v_2) \in E $, $ \min_{v_1} = c_i \min_{u_1} $ and $ \min_{v_2} = c_i \min_{u_2} $. Since $ \tau (v) = 3 $ we obtain that either (i) $ c_i \prec \lambda (u_1) $, or (ii) $ \lambda (u_1) = c_i $, $ \tau(u_1) = 3 $ and $ \psi (u_1) = \psi (I') - 1 =  \psi (I) - 1 $. In both cases, we conclude that $ u_1 $ is an element $ K $ of the suffix of $ \mathcal{A} $ whose largest element is $ I $, which by the inductive hypothesis has been correctly built, and so $ v_1 $ is added to an element of $ A_{c_i, t} $; $ v_1 $ is never removed from this element otherwise in $ I_{v_1} $ there would be a string smaller that $ \min_{v_1} $. As a consequence, it must also be $ u_2 \in K $, so by the inductive hypothesis $ \min_{u_1} = \min_{u_2} $ and we conclude $ \min_{v_1} = \min_{v_2} $.

        We are only left with proving that, if $ v_1 $ is neither in $ I' $, nor in the suffix of $ \mathcal{A} $ whose largest element is $ I $, and if $ v_2 \in I' $, then $ \min_{v_1} \prec \min_{v_2} $. The conclusion is immediate if $ \tau (v_i) \not = 3 $, so we can assume $ \tau (v_i) = 3 $. It cannot be $ c_i = \lambda (v_2) \prec \lambda (v_1) $ otherwise $ v_1 $ would be in the suffix of $ \mathcal{A} $ whose largest element is $ I $. Since $ \tau (v_i) = 3 $, it cannot be $ \lambda (v_1) = c_i $ and $ \psi(v_1) < \psi (I) $, otherwise again $ v_1 $ would be in the suffix of $ \mathcal{A} $ whose largest element is $ I $. As a consequence, it must $ \lambda (v_1) \preceq c_i $ and, if $ \lambda (v_1) = c_i $, then $ \psi (I') = \psi (I) \le \psi (v_1) $. If $ \lambda (v_1) \prec c_i $, or $ \lambda (v_1) = c_i  $ and $ \psi (I') < \psi (v_1) $, we immediately conclude $ \min_{v_1} \prec \min_{v_2} $. Hence, we can assume $ \lambda (v_1) = c_i $, $ \tau (v_1) = 3 $ and $ \psi (v_1) = \psi (I') $. Let $ u_1, u_2 \in V $ be such that $ (u_1, v_1), (u_2, v_2) \in E $, $ \min_{v_1} = c_i \min_{u_1} $ and $ \min_{v_2} = c_i \min_{u_2} $. As before, we must have already considered $ u_1 $ and $ u_2 $, but since $ v_1 $ is not in $ I' $, by construction it must be $ \min_{u_1} \prec \min_{u_2} $ and so we conclude $ \min_{v_1} \prec \min_{v_2} $.
        \item Suppose that $ \psi (I') = \psi (I) + 1 $. Arguing as we did above we infer that  all $ v \in V_{c_i, 3} $ such that $ \psi (v) = \psi(I) $ are already in some element of the suffix of $ \mathcal{A} $ whose largest element is $ I $. By using this information, the same proof of the previous case shows that $ I' $ is correct.
    \end{enumerate}
    \item Assume that $ i' = i $ and $ t' \le 2 $. Since we have already built the $ \mathcal{A}_{c, 2} $'s and the $ \mathcal{A}_{c, 1} $'s, we only have to prove that, if $ t = 3 $, then all $ v \in V_{c_i, 3} $ are already in some element of the suffix of $ \mathcal{A} $ whose largest element is $ I $. Assume for sake of contradiction that this is not true for some $ v \in V_{c_i, 3} $, and choose $ v $ such that $ \psi(v) $ is as small as possible. By arguing as before, we conclude (by the minimality of $ \psi (v) $) that $ v $ must be in some $ A_{c_i, 3} $; but since $ t' \le 2 $, we conclude that $ v $ must be in some element of the suffix of $ \mathcal{A} $ whose largest element is $ I $, a contradiction.
    \item Assume that $ i' = i - 1 $ and $ t' = 3 $. In particular, $ \psi (I') = 1 $. As in the previous case, we obtain that if $ t = 3 $, then all $ v \in V_{c_i, 3} $ are already in some element of the suffix of $ \mathcal{A} $ whose largest element is $ I $. Moreover, $ \mathcal{A}_{c_i, 2} $ and $ \mathcal{A}_{c_i, 1} $ must necessarily empty because we have already built them, and $ I' $ is not contained in any of them. First, let us prove that, if $ v_1, v_2 \in I' $, then $ \min_{v_1} = \min_{v_2} $. Let $ u_1, u_2 \in V $ be such that $ (u_1, v_1), (u_2, v_2) \in E $, $ \min_{v_1} = c_{i - 1} \min_{u_1} $ and $ \min_{v_2} = c_{i - 1} \min_{u_2} $. Since $ \tau (v) = 3 $ we obtain that either (i) $ c_{i - 1} \prec \lambda (u_1) $, or (ii) $ \lambda (u_1) = c_{i - 1} $, $ \tau (u_1) = 3 $ and $ \psi (u_1) = \psi (I') - 1 $. However, case (ii) cannot occur because $ \psi (I') = 1 $, so it must be $ c_{i - 1} \prec \lambda (u_1) $, and we conclude that $ u_1 $ is an element $ K $ of the suffix of $ \mathcal{A} $ whose largest element is $ I $. As before, we conclude that also $ u_2 $ is in $ K $, $ \min_{u_1} = \min_{u_2} $ and $ \min_{v_1} = \min_{v_2} $.

    We are only left with proving that, if $ v_1 $ is neither in $ I' $, nor in the suffix of $ \mathcal{A} $ whose largest element is $ I $, and if $ v_2 \in I' $, then $ \min_{v_1} \prec \min_{v_2} $. The conclusion is immediate if $ \tau (v_i) \not = 3 $, so we can assume $ \tau (v_i) = 3 $. It cannot be $ c_{i - 1} = \lambda (v_2) \prec \lambda (v_1) $ otherwise $ v_1 $ would be in the suffix of $ \mathcal{A} $ whose largest element is $ I $. It cannot be $ \lambda (v_1) = c_{i - 1} $ and $ \psi(v_1) < \psi (I') $, because $ \psi (I') = 1 $. As a consequence, it must hold $ \lambda (v_1) \preceq c_{i - 1} $ and, if $ \lambda (v_1) = c_{i - 1} $, then $ 1 = \psi (I') \le \psi (v_1) $. If $ \lambda (v_1) \prec c_{i - 1} $, or $ \lambda (v_1) = c_{i - 1}  $ and $ 1 = \psi (I') < \psi (v_1) $, we immediately conclude $ \min_{v_1} \prec \min_{v_2} $. Hence, we can assume $ \lambda (v_1) = c_{i - 1} $, $ \tau (v_1) = 3 $ and $ \psi (v_1) = \psi (I') = 1 $. Let $ u_1, u_2 \in V $ be such that $ (u_1, v_1), (u_2, v_2) \in E $, $ \min_{v_1} = c_{i - 1} \min_{u_1} $ and $ \min_{v_2} = c_{i - 1} \min_{u_2} $. As before, we must have already considered $ u_1 $ and $ u_2 $, but since $ v_1 $ is not in $ I' $, by construction it must be $ \min_{u_1} \prec \min_{u_2} $ and so we conclude $ \min_{v_1} \prec \min_{v_2} $.

    \item Assume that $ i' = i  - 1 $ and $ t' \le 2 $. As before, we obtain that if $ t = 3 $, then all $ v \in V_{c_i, 3} $ are already in some element of the suffix of $ \mathcal{A} $ whose largest element is $ I $. Moreover, $ \mathcal{A}_{c_i, 2} $ and $ \mathcal{A}_{c_i, 1} $ must necessarily be empty because we have already built them, and $ I' $ is not contained in any of them. Since we have already built $ \mathcal{A}_{c_{i - 1}, 2} $ and $ \mathcal{A}_{c_{i - 1}, 1} $, we only have to prove that $ \mathcal{A}_{c_{i - 1}, 3} $ is empty. If it were not empty, in particular there would exists $ v \in V_{c_{i - 1}, 3} $ with $ \psi (v) = 1 $. If $ u \in V $ is such that $ (u, v) \in E $ and $ \min_v = c_{i - 1} \min_u $, then we conclude that it must be $ c_{i - 1} = \lambda (v) \prec \lambda (u) $. Then, $ u $ must be in some element of the the suffix of $ \mathcal{A} $ whose largest element is $ I $, so $ v $ would be in some element of $ A_{c_{i  - 1}, 3} $; but this is a contradiction, because $ t' \le 2 $.
\end{enumerate}

Lastly, if after considering $ I $ in $ A_{c_i, t} $ there is no element $ I' $ to consider, we can check that every $ v \in V $ is in some element of the suffix of $ \mathcal{A} $ whose largest element is $ I $ (and so have built $ \mathcal{A} $). Indeed, assume for sake of contradiction that this is not true. Consider the set $ S $ of all nodes $ v $ that do not satisfy these properties (it must be $ \tau (v) = 3 $); let $ S' \subseteq S $ be the set of all nodes $ v $ in $ S $ for which $ \lambda (v) $ is maximal, and let $ S'' \subseteq S' $ be the set of all nodes $ v $ in $ S' $ for which $ \psi (v) $ is minimal. Pick any $ v \in S'' $, and let $ u \in V $ such that $ (u, v) \in E $ and $ \min_v = \lambda (v) \min_u $. As before, we obtain that $ u $ must be in some element of the suffix of $ \mathcal{A} $ whose largest element is $ I $, so $ v $ should be in some $ A_{c, 3}$ and there would be an element $ I' $ to consider, a contradiction.

\end{document}